\documentclass[
aps,
pra,
showpacs,
amsmath,
amssymb,
onecolumn,
floatfix,
longbibliography,
letterpaper,
lengthcheck,
superscriptaddress%
]{revtex4-2}

\usepackage{graphicx}
\usepackage{amsthm}
\usepackage{amsmath}
\usepackage{bm} %
\usepackage{physics}
\usepackage{color}
\usepackage{hyperref}
\usepackage[normalem]{ulem}
\usepackage[english]{babel}
\usepackage{lipsum}  

\usepackage{dsfont}
\usepackage[percent]{overpic}
\usepackage{todonotes}
\usepackage{graphicx,xcolor}
\usepackage{caption}
\usepackage{subcaption}
\newcommand {\diff} {\mathrm{d}}
\newcommand {\expec} [1] {\langle #1\rangle}
\newcommand{\id}{1\!\!\!1}
\renewcommand{\tr}{\operatorname{Tr}}
\newcommand{\bvec}[1]{\boldsymbol{#1}} %
\renewcommand{\abs}[1]{\lvert #1\rvert}
\newcommand{\cavg}[1]{\langle #1 \rangle}

\newcommand{\remove}[1]{\textcolor{blue}{\ifmmode\text{\sout{\ensuremath{#1}}}\else\sout{#1}\fi}}
\newcommand{\add}[1]{#1}

\begin{document}

\title{A Hierarchical Approach to Quantum Many-Body Systems in Non-Markovian Environments}

  \author{Kai M\"uller}
  \affiliation{Institut für Theoretische Physik, Technische Universität Dresden, D-01062 Dresden, Germany.}

  \author{Kimmo Luoma}
  \affiliation{Department of Physics and Astronomy, University of Turku, 20014 Turku, Finland.}

  \author{Christian Sch\"afer}
  \email[Electronic address:\;]{christian.schaefer@tuwien.ac.at}
  \affiliation{Institute of Applied Physics, TU Wien, Wiedner Hauptstrasse 8-10/134, Vienna, 1040, Austria}

\date{\today}

\begin{abstract}

Quantum many-body systems in cavities combine the rich physics of condensed matter systems or quantum chemistry with strong coupling to the surrounding electromagnetic field.
In these systems, large Hilbert spaces, many-body interactions and strong system-environment coupling are all fundamental, posing a significant barrier for established methods in quantum optics and condensed matter physics. 
Here we propose a novel method based on a combination of the Bogoliubov–Born–Green–Kirkwood–Yvon (BBGKY) hierarchy \cite{Bonitz} and the Hierarchical Equations of Motion (HEOM) \cite{Tanimura_1990} to achieve a rigorous description of open many-body systems in contact with structured photonic and phononic baths. 
We rationalize that this stacked hierarchy accounts for spin-squeezing and superradiant emission despite its applicability to arbitrarily many emitters.
The potential of BBGKY-HEOM is then demonstrated for many-body electronic systems embedded in host materials (e.g. molecules in organic crystals). 
We show that the impact of phononic coupling and charge noise can be as relevant as electronic correlation.
Our work establishes an accessible, yet rigorous, route between condensed matter and quantum optics, fostering the growth of a new domain at their interface. 
\end{abstract}

\date{\today}

\maketitle

\noindent
\section{Introduction} 
The ability to control many-body quantum systems with quantized light is imperative in modern quantum technologies, including photonics, quantum memories, quantum sensing, or quantum information.\cite{toninelli2021single,doi:10.1126/science.abb2823,Gambetta_2017,fleischhauer2002quantum,ohmanPredictive}
An idealized description of the fundamental building block, the quantum emitter, is challenged by disturbances, such as charge noise or phonon scattering, that hint at the inherent many-body character of each quantum system.
Strong light-matter coupling allows to control interactions over extended length scales\cite{coles2014b,feist2015,schafer2019modification,jarc2023cavity,schafer2023chiral}, holding promise to design materials~\cite{garcia2021manipulating,haugland2021intermolecular,latini2021ferroelectric,Schlawin2022Mar,Flores-Calderon2025Jan,Chakraborty2025Jun}, devices~\cite{leppala2024linear,abdelmagid2024identifying,lackner2024topologically}, or even chemical properties~\cite{schafer2021shining,ahn_herrera_simpkins_2022,doi:10.1021/jacs.3c02260,doi:10.1021/acs.jpclett.3c03506,mandal2023theoretical} non-intrinsically and on demand.
The many-body physics originating from electronic and vibrational structure, as well as their respective decoherence, is crucial for a comprehensive understanding and refinement of this control-strategy \cite{Schaefer_2024,sidler2024unraveling,lindoy2023quantum,siltanen2024incoherent,chan2021stable}.
Relevant experimental realizations include molecules or solids in micro- and plasmonic cavities, requiring a holistic treatment of electronic many-body physics coupled to various optical and phononic modes.\cite{Yang_2024,leppala2024linear}

Our primary challenge in this study is to address physical processes wherein many-body interactions, strong light-matter coupling, and strong system-environment coupling are all fundamental. 
Achieving this goal necessitates combining concepts from quantum optics and solid-state physics. Quantum optics often treats light-matter interaction non-perturbatively and offers many sophisticated techniques for dealing with strongly coupled environments \cite{HOPS,nuHOPS,Link_2023,TEMPO,Fowlwer-Wrigth_2022,uniTEMPO,ACE,TEDOPA,TEDOPA_algorithm,Pseudomode_Plenio,Pseudomode_Lin,ML-MCTDH,carmichael1999statistical,Davydov_review,Davydov_Frank,Tanimura_1990,Tanimura_2006}, but it typically simplifies the structure of the materials drastically.
Conversely, condensed matter or quantum chemistry approaches accurately describe the material system but oversimplify interactions with surrounding environments \cite{schafer2021making,haugland2021intermolecular,latini2021ferroelectric}. 
Here, we show how to bridge system-bath, quantum optical, and condensed matter approaches. We establish a framework for studying the effects of strongly coupled bosonic fields on many-body physics more efficiently than previously possible.

In order to deal with the strongly coupled environments we rely on the Hierarchical Equations Of Motion (HEOM)~\cite{Tanimura_1990,Tanimura_2006} approach. 
HEOM has been used successfully to solve a large variety of problems \cite{Tanimura2020,Fay2022Nov,Ma2012Jun,Kato2016Dec,Jin2008Jun,Ishizaki2009Jun,Hartle2013Dec,Chen_2015, Baetge_2021,Xu2026Jan}. It  lends itself to our cause particularly well, because (i) its approximation of the environment is completely independent of the system dimension, and (ii) it can be modified to incorporate nonlinear terms (see Sec.~\ref{sec:bbgkyheom}). 
However, its original implementation treats the system part exactly and thus fails for many-body systems due to the exponential scaling of the system dimension.

For the treatment of many-body systems wavefunction-based approaches such as tensor networks~\cite{Schollwock2011Jan} provide reliable results once convergence with respect to the bond dimension has been reached~\cite{Flannigan_2022,Daley_2014,Ke23,PhysRevLett.121.227401}.
\add{However, the computational cost of tensor network approaches often increases quickly over the course of a simulation, which, in many cases, limits their practical application to short timescales or one-dimensional systems.}
Non-equilibrium Green functions approaches \cite{Rao2023Feb} made recent progress after the successful integration of the generalized Kadanoff-Baym ansatz~\cite{PhysRevLett.124.076601,Tuovinen_2023,pavlyukh2025open} but a consistent treatment of complex structured baths remains challenging. And while molecular aggregates with finite range interactions can be very efficiently described with the help of non-Markovian quantum trajectories \cite{mesoHOPS,Gera2025Jun}, these methods will generally lose their scaling advantage for the all-to-all interactions mediated by the cavity.
An attractive approach with good scaling properties is to combine the treatment of the bath with a cluster expansion of the system density matrix \cite{Mu2026Mar} into single-, two-, three-, and $N$-body reduced density matrices and then truncate the resulting Bogoliubov–Born–Green–Kirkwood–Yvon (BBGKY) hierarchy at some low order~\cite{Bonitz,PhysRevResearch.5.033022}. 

Here, we combine BBGKY and HEOM to establish a stacked BBGKY-HEOM hierarchy that provides an efficient framework for the simulation of correlated many-body systems subject to structured baths.
Our method allows us to go past the paradigm of the two-level emitter and demonstrate that BBGKY-HEOM is able to capture many-body electronic dynamics in the Fermi-Hubbard model coupled to structured environments. 
BBGKY-HEOM provides accurate predictions as long as the onsite repulsion $U$, and with it the degree of electronic correlation remains moderate $U/J\lesssim 1$. 
In an accompanying letter \cite{letter_placeholder}, we apply our BBGKY-HEOM method to an experimentally relevant few-emitter lasing setup comprising multiple molecules inside a plasmonic cavity. Our results go beyond the state of the art and give new insights into how the vibrational spectrum influences the lasing process.
Especially interesting are resonant enhancements that can be linked to a relation between the drive strength and the plethora of vibrational modes supported by each molecule.

The remainder of this article is structured as follows.
Sec.~\ref{sec:theory} provides a short introduction to non-Markovian quantum systems and many-body theory before the derivation of the BBGKY-HEOM equations in Sec.~\ref{sec:bbgkyheom}.
We illustrate their performance, limitation, and associated physical intuition for various systems in Sec.~\ref{sec:results}.
After benchmarking our method for many-emitter quantum optics in Sec.~\ref{sec:manyemitter}, we shift our focus to many-body electronic systems in Sec.~\ref{sec:hubbard}, where we investigate the Fermi-Hubbard model coupled to a lossy cavity mode and subject to charge noise. This system is then embedded into a highly structured bath in Sec.~\ref{sec:hubbardwildbath} that represents phononic motion in organic crystals.
Sec.~\ref{sec:conclusion} finally concludes our study and provides an outlook into the near future.

\section{Theory}\label{sec:theory}

A material comprising electrons and nuclei moving at non-relativistic velocities is described by Schrödinger's equation. Its interplay with electromagnetic fields necessitates the consideration of the (quantized) normal modes of Maxwell's equations. This combined system can be described using the minimal coupling Hamiltonian. In Coulomb gauge,
\begin{align}\label{eq:coulomb_full}
\begin{split}
    \hat{H}_{mc} &= \sum_{i=1}^{N_e+N_n} \frac{1}{2m_i} \big[ -i\hbar \boldsymbol\nabla_i - q_i \hat{\textbf{A}}(\textbf{r}_i) \big]^2
    + \hat{H}_\parallel\\
    &+ \frac{\varepsilon_0}{2}\int dr^3 \big[ \hat{\textbf{E}}_\perp(\textbf{r})^2 + \hat{\textbf{B}}(\textbf{r})^2 \big]
\end{split}
\end{align}
for $N_e$ electrons and $N_n$ nuclei with charge $q_i$ and mass $m_i$. %
We will use atomic units in the following.
Eq.~\eqref{eq:coulomb_full} splits the transversal degrees from the longitudinal and instantaneous Coulombic interactions $H_\parallel$.
The latter includes electronic and nuclear interactions which are initially treated in first quantization 
\begin{align}
\begin{split}
    \hat{H}_\parallel &= \hat{V}_{ee} + \hat{V}_{en} + \hat{V}_{nn}; \quad 
    \hat{V}_{ee} = \sum_{i< j}^{N_e} \frac{1}{\vert \textbf{r}_i - \textbf{r}_j \vert}\\
    \hat{V}_{en} &= \sum_{i,j}^{N_e,N_n} \frac{-q_i}{\vert \textbf{r}_i - \textbf{R}_j \vert}; \quad 
    \hat{V}_{nn} = \sum_{i< j}^{N_n} \frac{q_i q_j}{\vert \textbf{R}_i - \textbf{R}_j \vert}.
\end{split}
\end{align}
Their interplay results in rich many-body physics that gives rise to electronic, vibronic, and phononic structure and, unfortunately, to an exponentially increasing complexity.
The open quantum systems approach now aims to identify a small subsystem of interest that we can treat explicitly (the "system") while absorbing the remainder into  environments (the "baths") \cite{OpenQuantumSystems}. The first step is to simplify the Hamiltonian to a form that contains only linear interactions between system and bath(s)
\begin{align}
    \hat{H}_{tot} = \hat{H}_{sys} + \hat{H}_B + \hat{H}_{sys,B}.
\end{align}

The gauge freedom of the electromagnetic modes provides various different paths for this. %
An intuitive approach is separate the bilinear term $\sum_i-i\hbar q_i/m_i \hat{\textbf{A}}(\textbf{r}_i)\cdot \boldsymbol\nabla_i$ from the modified momentum $\sum_{i=1}^{N_e+N_n} \frac{1}{2m_i} \big[ -i\hbar \boldsymbol\nabla_i - q_i \hat{\textbf{A}}(\textbf{r}_i) \big]^2$ introduced in Eq.~\eqref{eq:coulomb_full} as coupling operator and absorb the quadratic diamagnetic term {$\sum_{i=1}^{N_e+N_n} \frac{q_i^2}{2m_i} \hat{\textbf{A}}(\textbf{r}_i)^2$} into dressed photonic operators~\cite{schafer2019relevance,Nazir2022}.
A second direction, often more promising for finite systems, is to shift into the multi-polar gauge (detailed in App.~\ref{app:multi}). 
The combination with the long-wavelength approximation{, which demands that the wave length of the field is much larger than the extend of the matter system,} then provides a simple bilinear coupling between {electric} dipole moment and displacement field. 
In addition, self-polarization terms modify the dynamics of the system and the redefinition of the optical creation and annihilation operators introduce the need to reconsider the interpretation of system and bath.
Vibrational and phononic interaction follow similarly. 
The influence of nuclear motion on the electron-nuclear interaction can be expanded around the equilibrium configuration
\begin{align}
    \hat{V}_{en} \approx \hat{V}_{en}\vert_{0} + \sum_j \delta \textbf{R}_j \cdot \boldsymbol\nabla_j\hat{V}_{en}\vert_{0} + ...
\end{align}
and truncated at harmonic order, resulting in a bilinearly coupled electron-phonon or electron-vibration coupling that depends on the electronic configuration. Note that also here second-order (known as Debye-Waller) corrections appear that require to carefully scrutinize the separation between system and bath \cite{miglio2020predominance}.
The following derivations and demonstrations are agnostic to such subtleties, as we will focus on the development and illustration of the BBGKY-HEOM methodology. A thorough derivation of system-bath couplings for a specific physical realization is evidently essential for meaningful predictions but it is not the focus of this manuscript.

\subsection{Many-body Open Quantum Systems beyond the Markovian Limit}\label{ssec:introManyBodyOQS}

Following the steps outlined above, both the phonon and the (transversal) photon sector of $\hat{H}_{mc}$ can be described as a collection of quantum harmonic oscillators that couple linearly to the electronic system. This description of an environment is frequently the case in quantum optics~\cite{IntroQuantumOptics}, or when the system is coupled weakly to a large number of degrees of freedom where central limit type of arguments yield an approximately Gaussian response~\cite{makri1999linear}. 
We label these distinct environments (e.g. a photonic and a phononic bath) with $k$, such that the bosonic annihilation (creation) operators belonging to mode $\lambda$ of environment $k$ are given by $a_\lambda^k$ ($a_\lambda^{k\dagger}$).
The Hamiltonian corresponding to the total state $\rho_{tot}$ of system and environments is then
\begin{align}
    \label{eq:Htot}
    H =& H_{sys} + \sum_{k,\lambda} g_\lambda^k(L^k a_\lambda^{k,\dagger} + L^{k\dagger} a^k_\lambda) +\sum_{k,\lambda} \omega^k_\lambda a_\lambda^{k\dagger} a^k_\lambda.
\end{align}
Here %
$L^k$ is the system operator that describes the coupling to the non-Markovian environment $k$ and $\omega_\lambda^k$, $g_\lambda^k$ are the frequencies and coupling strengths of the respective modes.
The environments are characterized by their spectral density $J^k(\omega) = \sum_\lambda \abs{g^k_\lambda}^2 \delta(\omega-\omega^k_\lambda)$ or equivalently by their bath correlation function at temperature T $\alpha^k(\tau) = {\int_0^\infty J_k(\omega)\left(\coth{(\omega/(2T))}\cos{(\omega\tau)} -i\sin{(\omega\tau})\right) \diff{\omega}}$.
When a description of the environment in terms of harmonic oscillators is suitable and the initial state of system and environment is a product state, the HEOM approach yields a numerically exact description for the evolution of the open quantum system, even for strong coupling \cite{Tanimura_1990,Tanimura_2006,Tanimura2020}. To use HEOM, the bath correlation function needs to be fitted with exponential functions~\cite{Liu_2014} according to
\begin{equation}\label{eq:fitExp}
    \alpha^k(\tau) \approx \sum_j^{N_{exp}}G_j^k \exp\left(-{W_j^k}\tau\right),
\end{equation}
with $G_j^k,\;{W_j^k} \in \mathbb{C}$.
The resulting hierarchy couples the physical density matrix to auxiliary density matrices, and the number of auxiliary density matrices depends on the number of exponentials $N_{exp}$ as well as the system-environment coupling strength. {Due to the scaling with $N_{exp}$ an efficient fitting algorithm can significantly increase the efficiency of HEOM. Several algorithms are available for this task \cite{expFittingOverview}, both in the time \cite{PronyMethod} and the frequency domain \cite{AAA_algorithm, Xu2022Nov}.}\\
In fact, HEOM type of techniques can be even extended to cases where the environment is anharmonic~\cite{Hsieh_2018_a,Hsieh_2018_b}.
In this work, we focus on the case of harmonic baths, either corresponding to optical modes confined in an imperfect cavity or a phonon bath into which a quantum emitter is embedded. 
The influence functional of the bath can then be computed exactly under the harmonic assumption by using the path integral formalism~\cite{weiss2008quantum}. 
HEOM follows from this influence functional~\cite{Tanimura_2006} and we provide a sketch of the derivation for the open system in App.~\ref{app:HEOM}.\\

Following the previous motivation, we want to treat a many-body system that couples to the environments. 
The $N$-body Hamiltonian can be decomposed into a single particle Hamiltonian $H_i$ and two-particle interactions $V_{ij}$
\begin{align}\label{eq:Hsys}
    \begin{split}
        H_{sys} =& \sum_{i=1}^N H_i + \sum_{i < j}^N V_{ij}.
    \end{split}
\end{align}
{Additional (local) Markovian dissipative processes of the system are captured by $\mathcal{L}_{diss}$.}
The evolution of the total state $\rho_{tot}$ of system and environments is then described by
\begin{align}
    \label{eq:fullEvolution}
    \dot{\rho}_{tot} =& -i[H, \rho_{tot}] + \mathcal{L}_{diss}(\rho_{tot}).
\end{align}
Eq.~\eqref{eq:fullEvolution} together with Eq.~\eqref{eq:Htot}\eqref{eq:Hsys} provide the general form for systems discussed in this work. %
HEOM retains some information about the bath, allowing one to access e.g. photonic correlation functions. %

For the coupling operators $L_k$ we distinguish two cases, both illustrated in the following. First, all particles couple to the same environment with equal strength and the coupling operators take the form
\begin{equation}\label{eq:identicalCoupling}
    L^k = \sum_{i=1}^N L^k_i.
\end{equation}
Second, each particle couples to its own local environment, however, all local environments are identical. The latter situation arises, for example, in a collection of molecules, each with its local vibrational bath \cite{letter_placeholder}. The coupling operators are then denoted as $S_k$, and act only on the local Hilbert space of particle $k$.
In both cases the evolution equation \eqref{eq:Htot} is invariant under particle exchange, which we will leverage in the following.
Such assumptions are natural for indistinguishable particles but one can relax this condition if one would like to describe ensembles of inhomogeneously broadened emitters.

Despite the restrictions employed above, the resulting system-bath Hamiltonian applies to a vast range of systems critical in fields such as quantum optics, quantum impurities in solid hosts, and strongly correlated materials in cavities. 

\subsection{Combining HEOM with BBGKY}\label{sec:bbgkyheom}

Let us now tackle the exponential growth of the Hilbert space by stacking HEOM {\cite{Tanimura_1990, Tanimura2020}} and BBGKY {\cite{Bonitz}} hierarchy.
We focus here on a single environment for the sake of simplicity, but note that the extension to multiple environments is trivial and will be illustrated in Sec.~\ref{sec:hubbardwildbath}.
The HEOM equation accounts for the non-Markovianity of the coupled environment with an infinite set of coupled equations for the matrices $\rho^{(\bvec{n},\bvec{m})}$, whose dimension is equal to the dimension of the system Hilbert space. 
Here, $\rho^{(\bvec{0},\bvec{0})} = \tr_{B}(\rho_{tot}) =: \rho_{sys}$ is the physical density matrix of the system and $\bvec{m},\bvec{n}$ are vector indices {with dimension $N_{exp}$; $\bvec{n},\bvec{m} \in \mathbb{N}^{N_{exp}}$.} {They} label the auxiliary density matrices.  
{A physical intuition for these auxiliary matrices can be gained from related pseudomode models, as shown in Ref.~\cite{Muller2026Apr}. In practice, the number hierarchy indices $\bvec n,\,\bvec m$ must remain finite and thus an increasing number is included until a convergence is reached.} The hierarchy {can then be} truncated at the sufficiently large index. {The number of required indices generally scales with the system-bath coupling strength, the memory time of the bath and the number of exponential terms $N_{exp}$ used in Eq.~\eqref{eq:fitExp}.} A Markovian bath corresponds to a truncation at first order according to $\rho^{(\bvec{e}_k,\bvec{0})} \propto L\rho^{(\bvec{0},\bvec{0})}$, $\rho^{(\bvec{0},\bvec{e}_k)} \propto \rho^{(\bvec{0},\bvec{0})}L^\dagger$, where $\bvec{e}_k$ denotes the unit vector in direction $k$. 
The evolution equation for the auxiliary density matrices (including the physical reduced density matrix $\rho^{(\bvec{0},\bvec{0})}$) for a single environment with coupling operator $L$ reads
\begin{align}\label{eq:HEOM}
        \dot{\rho}^{(\bvec{n},\bvec{m})}&=-i[H_{sys}, \rho^{(\bvec{n},\bvec{m})}] - \left({\bvec{W}}\cdot\bvec{n}+{\bvec{W}^*}\cdot\bvec{m}\right)\rho^{(\bvec{n},\bvec{m})}\notag\\
        &+\sum_{j=1}^{N_{exp}}\bigg(G_jn_jL\rho^{(\bvec{n}-\bvec{e}_j,\bvec{m})} + G_j^*m_j\rho^{(\bvec{n},\bvec{m}-\bvec{e}_j)}L^{\dagger}\notag\\
        &\quad\quad\quad+\left[\rho^{(\bvec{n}+\bvec{e}_j,\bvec{m})}, L^{\dagger}\right] + \left[L, \rho^{(\bvec{n},\bvec{m}+\bvec{e}_j)}\right]\bigg).
\end{align}
{Here and throughout the manuscript b}old symbols indicate the vector character of a variable. The complex coefficients $G_j, {W_j}$ can be obtained from a fit according to Eq.~\eqref{eq:fitExp} or inferred from scattering theory \cite{Lentrodt20}. 
The application to many-body systems now requires us to identify a path to limit the dimensionality of the descriptor, so far $\rho^{(\bvec{n},\bvec{m})}$, to manageable levels.

An intuitive way to accomplish this is presented by the BBGKY hierarchy which is build around the reduced density matrices (RDM)
\begin{equation}\label{eq:RDM}
    \begin{split}
        F_{123}^{(\bvec{n}, \bvec{m})} =& N(N-1)(N-2)\tr_{4,5,...,N}(\rho^{(\bvec{n}, \bvec{m})}),\\
        F_{12}^{(\bvec{n}, \bvec{m})} =& N(N-1)\tr_{3,4,...,N}(\rho^{(\bvec{n}, \bvec{m})}),\\
        F_{1}^{(\bvec{n}, \bvec{m})} =& N\tr_{2,3,...,N}(\rho^{(\bvec{n}, \bvec{m})}).
    \end{split}
\end{equation}
Here $\tr_{i, i+1, ...,N}$ denotes the trace over all particles with index larger than $i$. The reduced density matrices are normalized to account for the particle number under contraction. Note that since we are restricting ourselves to permutationally symmetric systems, the order of indexing is irrelevant. {With this permutational symmetry in mind we will stick to the convention that subscripts only denote which particle(s) an operator is acting on. The different RDMs in Eq.~\eqref{eq:RDM} are distinguished by the number of subscripts. For example $F_{2}^{(\bvec{n}, \bvec{m})} = N\id\otimes\tr_{2,3,...,N}(\rho^{(\bvec{n}, \bvec{m})})$ and $F_{23}^{(\bvec{n}, \bvec{m})} = N\add{(N-1)}\id\otimes\tr_{3,4,...,N}(\rho^{(\bvec{n}, \bvec{m})})$. Furthermore, we use the same name irrespective of the total dimension of an operator, e.g., $\hat{O}_2$ denotes both $\id\otimes\hat{O}\otimes\id\otimes ...\otimes\id$ and $\id\otimes\hat{O}$. The intended dimension becomes clear from the context.}

Our goal in the following is to derive a closed hierarchy of equations for the reduced density matrix $F_{12}$, such that the (auxiliary) density matrices in Eq.~\eqref{eq:HEOM} only scale with the square of the dimension of the single-particle Hilbert space. 
From Eq.~\eqref{eq:HEOM} and with {$H_{sys}$,} $L$ given in the form of Eq.~{\eqref{eq:Hsys} and }\eqref{eq:identicalCoupling} {respectively, }it follows that (see App.~\ref{app:derivation} for derivation and App.~\ref{app:localBaths} for local baths) %
\begin{widetext}
\begin{align}\label{eq:reducedHEOM}
        \dot{F}_{12}^{(\bvec{n},\bvec{m})}&=-i[H_1 + H_2 \add{+V_{12}}, F_{12}^{(\bvec{n},\bvec{m})}] -i \tr_3\left([V_{13} + V_{23}, F_{123}^{(\bvec{n},\bvec{m})}]\right)- \left({\bvec{W}}\cdot\bvec{n}+{\bvec{W}^*}\cdot\bvec{m}\right)F_{12}^{(\bvec{n},\bvec{m})}\notag\\
        &+\sum_{k=1}^M\bigg(G_k\left(n_k(L_1+L_2)F_{12}^{(\bvec{n}-\bvec{e}_k,\bvec{m})} +n_k\tr_3(L_3F_{123}^{(\bvec{n}-\bvec{e}_k,\bvec{m})})\right) +G_k^*\left(m_kF_{12}^{(\bvec{n},\bvec{m}-\bvec{e}_k)}(L_1^{\dagger} + L_2^\dagger) + m_k\tr_3(L_3^\dagger F_{123}^{(\bvec{n},\bvec{m}-\bvec{e}_k)})\right)\notag\\
        &\quad\quad+\left[F_{12}^{(\bvec{n}+\bvec{e}_k,\bvec{m})}, L_1^{\dagger} + L_2^\dagger\right] + \left[L_1 + L_2, F_{12}^{(\bvec{n},\bvec{m}+\bvec{e}_k)}\right]\bigg).
\end{align}
\end{widetext}
For any interacting many-body system, the two-body RDM $F_{12}$ will couple to the three-body RDM $F_{123}$, which in turn will couple to the four-body RDM and so on.
In order to close this hierarchy we express the three-body RDM in terms of the one- and two-body RDM 
\begin{equation}\label{eq:reconstruction}
    F_{123} \approx \widetilde{F}_{123}(F_{12}, F_1),
\end{equation}
where $\widetilde{F}_{123}$ should be understood as a function that approximates the true three-body RDM. {This step is the core approximation performed in the derivation of BBGKY-HEOM.} We demonstrate in App.~\ref{app:derivation} how {it} can be {derived} by reformulating the hierarchy of equations for the density matrices $\rho_{sys}^{(\bvec{n},\bvec{m})}$ as an equation for a \textit{single} operator {acting on} an extended Hilbert space $\mathcal{H}_{ext} = \mathcal{H}_{sys}\otimes\mathcal{H}_{aux}$ and subsequently neglecting {non-Gaussian} three-body correlations within the system. 
However, we keep correlations between two particles in the system and the auxiliary degree of freedom that captures the influence of the environment.
{Note that, unlike the widely used cumulant expansion \cite{cumulants,quantum_cumulants_jl}, our approximation still allows for non-Gaussian single-particle states, since only non-Gaussian correlations \emph{between} particles are neglected. This small but crucial difference enables us, for example, to treat non-Gaussian vibrational states that are relevant for polaritonic chemistry \cite{Schwengelbeck2026Apr}.}\\
With these approximations we arrive at the expression below for the three-body RDM, where we have left out indices $(\bvec{0}, \bvec{0})$ (which correspond to the physical RDMs) for brevity
\begin{align}\label{eq:F123}
        \widetilde{F}^{(\bvec{n}, \bvec{m})}_{123} &= 4\frac{(N-1)(N-2)}{N^3}\tr(F_1^{(\bvec{n},\bvec{m})}) F_1 F_2 F_3\notag\\
        &+\frac{N-2}{N}\left(F_{12} F_3^{(\bvec{n},\bvec{m})} + F_2^{(\bvec{n},\bvec{m})}F_{13} + F_1^{(\bvec{n},\bvec{m})} F_{23}\right) \notag\\
        &+ \frac{N-2}{N}\left(F_{1} F_{23}^{(\bvec{n},\bvec{m})} + F_{13}^{(\bvec{n},\bvec{m})}F_{2} + F_{12}^{(\bvec{n},\bvec{m})} F_{3}\right) \notag\\
        &- \frac{N-2}{N^2}\tr(F_1^{(\bvec{n},\bvec{m})})\left(F_{12}F_3 + F_{13}F_2 + F_1F_{23}\right)\notag\\
        &-2\frac{(N-2)(N-1)}{N^2}\Big(F_1^{(\bvec{n},\bvec{m})} F_2F_3 \notag\\
        &\quad+ F_1F_2^{(\bvec{n},\bvec{m})}F_3 + F_1F_2F_3^{(\bvec{n},\bvec{m})}\Big).
\end{align}

{As a reminder, we note that, following our convention, terms such as $F_{12} F_3^{(\bvec{n},\bvec{m})}$ and $F_1^{(\bvec{n},\bvec{m})}F_{23}$ differ only in the tensor product order, since $F_1^{(\bvec{n},\bvec{m})}$ and $F_3^{(\bvec{n},\bvec{m})}$ (as well as $F_{12}$ and $F_{23}$) correspond to identical operators that are, however, acting on different particles.}
Inserting the above approximation into Eq.~\eqref{eq:reducedHEOM} provides the BBGKY-HEOM equation up to three-body correlations -- the central tool of this work. As desired, it represents a closed hierarchy for the two-particle RDM that can account for interactions within the system and simultaneously includes the influence of a generic non-Markovian environment.\\
$\widetilde{F}^{(\bvec{m}, \bvec{n})}_{123}$ is invariant under particle exchange, but does not preserve the (anti-)symmetry of the wavefunctions as required for (fermionic) bosonic particles. 
{Reconstructing an (anti-)symmetric 3-RDM is more complicated, since not all antisymmetric three-particle density matrices can be obtained as a \emph{reduced} state of an anti-symmetric $N$-particle density matrix. To guarantee this, a set of additional representability conditions \cite{N_representability_conditions} needs to be satisfied. 
In practice, these conditions are hard to check and a naive projection of the reconstructed 3-RDM into the (anti-) symmetric subspace can lead to instabilities. 
For a practical solution we follow here the approach in Refs.~\cite{Lackner,Lackner2015Feb,PhysRevResearch.5.033022}, where it was shown that enforcing contraction consistency of the reconstructed 3-RDM $\operatorname{Tr}_3(F_{123}) = \add{(N-2)}F_{12}$ and implementing a purification scheme can significantly increase the stability and quality of the BBGKY calculations. We describe our implementation of these two procedures in detail in Appendix~\ref{app:purification}. In particular the reconstructed anti-symmetric 3-RDM is given in Eq.~\eqref{eq:antiSymmetricReconstruction}.}

\section{Results}\label{sec:results}
In the following, we demonstrate the applicability, strengths, and limitations of BBGKY-HEOM {on typical models from quantum optics and condensed matter physics}.
Benchmarks via the driven Tavis-Cummings and Cavity-Fermi-Hubbard models serve to demonstrate the excellent performance of BBGKY-HEOM.
We then simulate system-bath competition in a Cavity-Fermi-Hubbard model embedded in a structured phonon bath, representing optical processes in organic crystals.

\begin{figure*}[t]
    {
        \centering
        \begin{subfigure}[b]{0.59\textwidth}
             \centering
             \includegraphics[width=\textwidth]{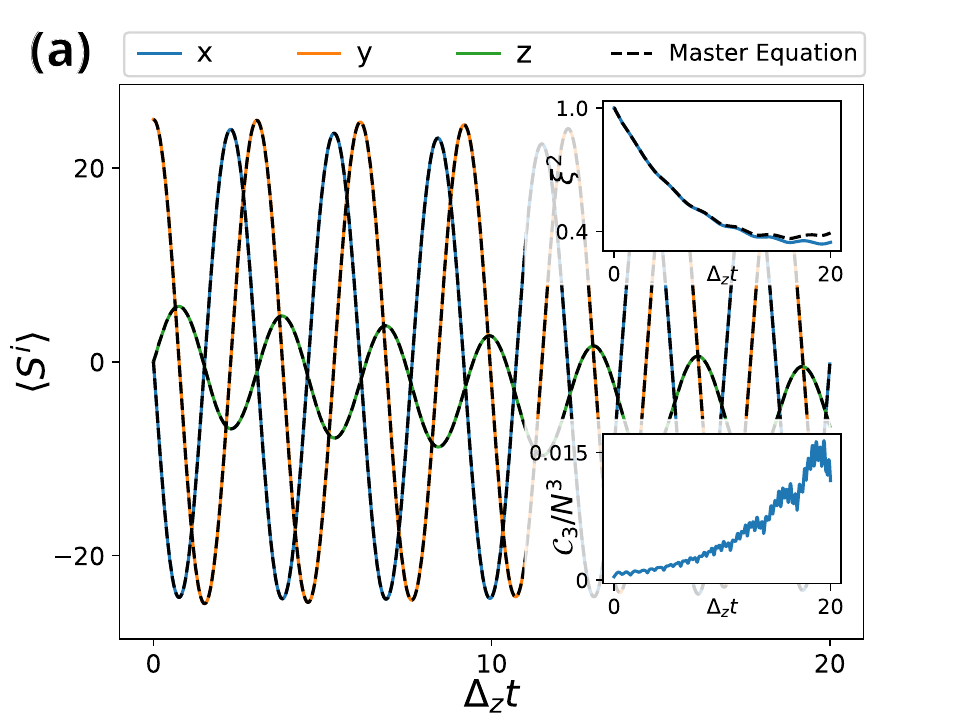}
         \end{subfigure}
         \begin{subfigure}[b]{0.39\textwidth}
             \centering
             \includegraphics[width=\textwidth]{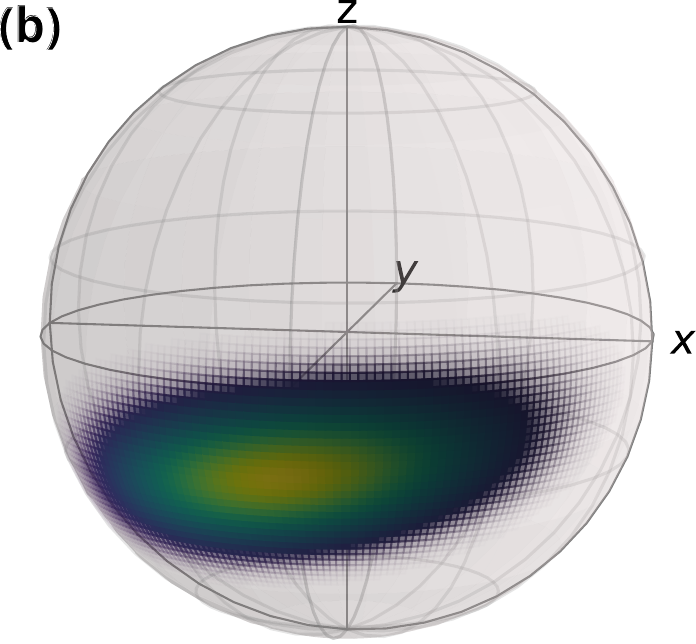}
         \end{subfigure}
         \begin{subfigure}[b]{0.49\textwidth}
             \centering
             \includegraphics[width=\textwidth]{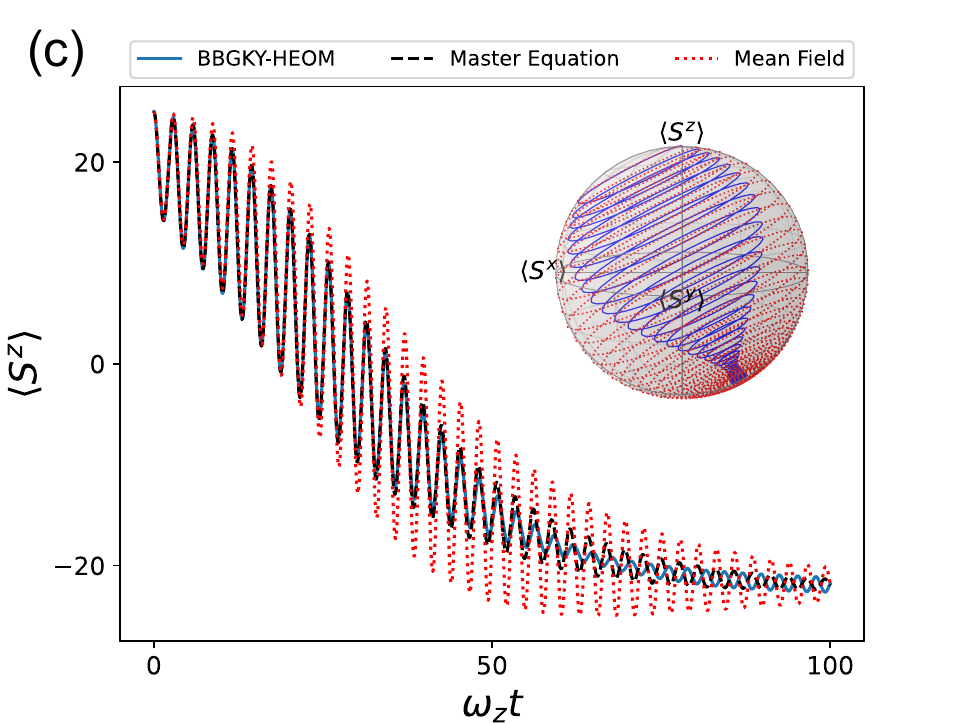}
         \end{subfigure}
         \begin{subfigure}[b]{0.49\textwidth}
             \centering
             \includegraphics[width=\textwidth]{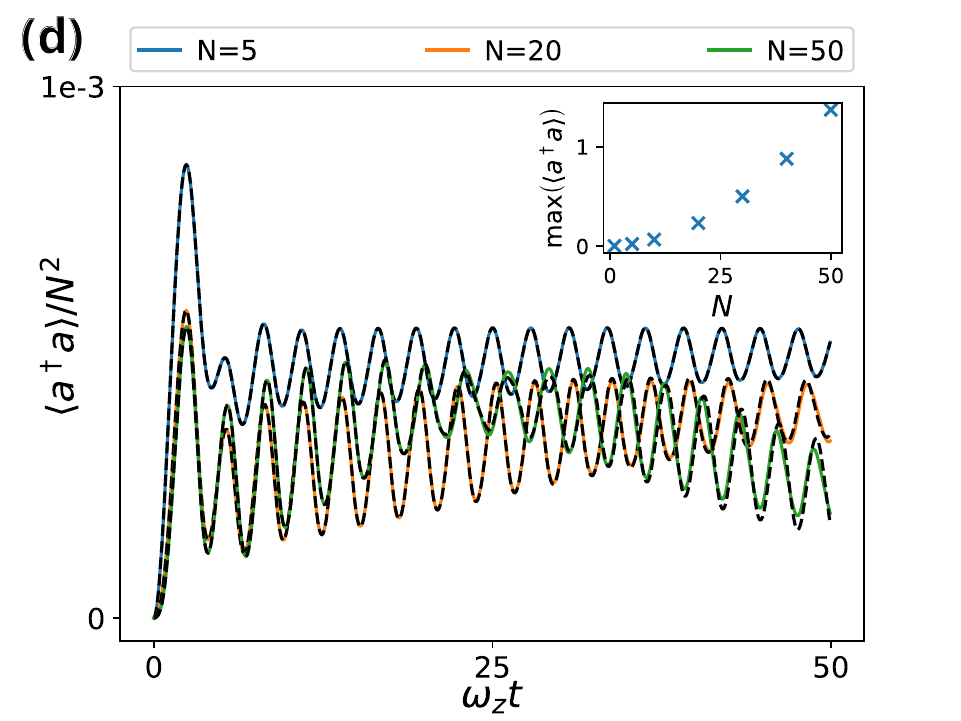}
         \end{subfigure}
    }
    \caption{
    \textbf{Squeezing and superradiance in the driven Tavis-Cummings model}: 
    (a) Spin dynamics for $N=50$, $ g\sqrt{N}/\kappa = 0.5$, $\Delta_z/\kappa=0.2$, $\Omega/\kappa = 0.05$, $\Delta_c/\kappa = 10$.  
    (b) Corresponding spin-squeezing (see text).
    (c) Superradiant emission from fully excited state. Parameters adjusted to $\Delta_z/\kappa=2$, ${\Omega}/\kappa=1$, $\Delta_c/\kappa=2.4$, $g/\kappa = 0.1$.
    (d) Photon-occupation normalized by $N^2$ over time and maximum photon number in burst (not normalized) for different N obtained with BBGKY-HEOM.
    BBGKY-HEOM provides reliable predictions and only begins to deviate slightly at long times when explicit 3-body correlations accumulate to noticeable values.
    }
    \label{fig:drivenTC}
\end{figure*}

\subsection{Driven-dissipative many-emitter systems}\label{sec:manyemitter}
A collection of emitters that share a common mode can interact with it collectively, potentially resulting in burst-like correlated emission known as superradiance and non-classical squeezed spin states.
Following a series of simplifications for matter, cavity, and light-matter coupling, {these systems can be described with} the widely used Dicke Hamiltonian~\cite{IntroQuantumOptics}
\begin{equation}
    H_{Dicke} = \omega_a \sum_i \sigma_i^z + \omega_c a^\dagger a + g\sum_i \sigma_x^i(a^\dagger + a).
\end{equation}
The Pauli matrices $\sigma_i^{x,y,z}$ represent the SU(2) algebra of an idealized two-level emitter while $a^\dagger$($a$) are the creation (annihilation) operators of the cavity mode.
The collection of emitters can be brought out of equilibrium by coherently driving the system $H_{drive} = 2\Omega \cos{(\omega_d t)}\sum_i \sigma_i^x$.
A unitary transformation moves us into a frame that rotates with the driving frequency $\omega_d$.
As long as the drive is close to resonance, i.e., $\omega_d \gg \Delta_z, \Delta_{{c}}, g,\Omega $ with the detunings $\Delta_z = \omega_a - \omega_d$, $\Delta_c = \omega_c - \omega_d$, we can safely discard counter-rotating terms $e^{i(\omega_d+\omega_{a/c})}$ that oscillate too fast to influence the dynamics of the system.
We obtain the paradigmatic driven Tavis-Cummings Hamiltonian
\begin{equation}\label{eq:drivenDicke}
\begin{split}    
    H_{DTC} =& \Delta_z \sum_i \sigma_i^z + \Omega \sum_i \sigma_i^x + \Delta_c a^\dagger a\\ 
    +& {g}\sum_i \left(\sigma_i^+ a + \sigma_i^- a^\dagger\right).
\end{split}
\end{equation}
We will consider the emitters or spins in the following as our system of interest and the cavity as an environment that dissipates energy into free space at a rate $\kappa$. 
{For the sake of simplicity we restrict our example to a parameter regime, where this decay can be described by a GKSL master equation with Lindblad operator $a$ \cite{Scala2007, OpenQuantumSystems, SciPostPhysLectNotes.129}, such that}
\begin{equation}\label{eq:dampedDrivenDicke}
    \begin{split}
        {\dot{\rho} =}& {-i[H_{DTC},\rho] + \kappa\mathcal{D}[a](\rho),}\\
         {\mathcal{D}[a](\rho) =}& {\left(\add{2}a\rho a^\dagger -\left(a^\dagger a\rho + \rho a^\dagger a\right)\right).}
    \end{split}
\end{equation}
{In this case} the bath correlation function is readily obtained, for example, from the Heisenberg-Langevin equations \cite{Muller2026Apr, OpenQuantumSystems}, as
\begin{equation}\label{eq:alpha_cav}
    \alpha(\tau) = g^2\expec{a(\tau)a^{\add{\dagger}}(0)} = g^2e^{(-i\Delta_c - \kappa)\tau}.
\end{equation}
For this simple example, therefore, no fitting procedure is necessary and we can easily use Eq.~\eqref{eq:reducedHEOM} with $M=1$ and $G=g^2, W=i\Delta_c + \kappa$. 
Using the truncation of the many-body BBGKY hierarchy {in Eq.~\eqref{eq:reconstruction}}, we then evolve a hierarchy of 2-particle RDMs, here $4\cross 4$ matrices, augmented with the HEOM indices $(\textbf{m},\textbf{n})$. 
The computational complexity is therefore \textit{independent} of the number of emitters. 

Given the necessary computational resources, we could have just as well decided to directly solve the entire system of $N$-emitters plus lossy cavity using the Lindblad master equation {\eqref{eq:dampedDrivenDicke}}.
This exact solution {of Eq.~\eqref{eq:dampedDrivenDicke}} will be used to estimate the quality of the BBGKY-HEOM and can be obtained in a collective spin picture $S^k = \sum_i \sigma_i^k{/2}$, $k\in\{x,y,z,+,-\}$. 
The $2^N$ dimensional system Hilbert space can then be restricted to a basis of $N+1$ permutationally symmetric Dicke states $\{ \ket{N/2, m}, m= -N/2, -N/2 +1,\dots,N/2\}$
which results in a linear increase of the relevant basis size with $N$ and therefore a quadratic increase of the Liouville space. 
The collective spin-picture fails as soon as individual emitter-emitter interactions need to be included, resulting in an exponentially growing Hilbert space that prohibits exact solutions for systems larger than a handful of emitters.
However, the computational cost for BBGKY-HEOM is unaffected by this change and remains system-size independent. 
{Note, that e}ven if there is no direct interaction in this model, the spins still interact indirectly via the cavity mode.\\

Fig.~\ref{fig:drivenTC} (a) illustrates the system dynamics of Eq.~\eqref{eq:dampedDrivenDicke} for $N = 50$ emitters strongly coupled to the cavity $ \sqrt{N}g/\kappa = 0.5$ in comparison to the exact solution of the master equation.
The spin components $\cavg{S^{i}}$ are accurately predicted within the simulation window, despite the non-classical dynamics of the system.
Shown in the upper inset is the spin squeezing \cite{Kitagawa1993} (see App.~\ref{app:spinsqu}) which is exhibiting values smaller than unity -- a sufficient condition for entanglement \cite{Sorensen2001Jan}.
We observe small deviations at long times which can be understood by investigating the slow increase in the explicit 3-body correlations (lower inset), i.e., the term that is discarded in the currently chosen truncation.
The spin-squeezing is further illustrated by the spin-Q function $Q=\bra{\theta, \phi}\rho\ket{\theta, \phi}$ in the upper right subplot (b), with $\ket{\theta, \phi}$ the spin coherent state. Mean-field theory will fail in predicting any of the spin-squeezing effects.\\

Let us fully excite all emitters and illustrate the resulting superradiant emission in Fig.~\ref{fig:drivenTC} (c) (parameters adjusted, see caption).
BBGKY-HEOM is accurate, with small deviations to the exact solution only at long times near the fully de-excited steady-state.
While mean-field theory {captures the overall dynamics qualitatively}, it {performs significantly worse than BBGKY-HEOM at all times \footnote{The mean-field equations were obtained by writing the equations of motion for the expectation values and factorizing averages of operator products according to $\cavg{XY+YX}/2 = \cavg{X}\cavg{Y}$.}}.
The inset projects the dynamics on a sphere. The failure of mean field originates from restricting the dynamics to the surface of the sphere, whereas the correct dynamics traverses the Bloch ball diagonally.
The sudden increase in the photon number originating from the burst-like emission of $N$ correlated emitter is presented in Fig.~\ref{fig:drivenTC} (d), demonstrating again accurate predictions by BBGKY-HEOM.
Continuous driving results in a continuous replenishment of the cavity mode.
The maximum photon-number (inset) is correctly described in its quadratic increase with the emitter number $N$.
In conclusion, BBGKY-HEOM provides accurate predictions for matter and cavity observables at constant cost, i.e., for arbitrary many emitter and with the option to account for direct interaction.

\subsection{Many-body electronic systems coupled to non-Markovian optical environments}\label{sec:hubbard}

\begin{figure*}[t]
    \begin{minipage}{\textwidth}
        \centering
        \begin{subfigure}[b]{0.47\textwidth}
             \centering
             \includegraphics[width=\textwidth]{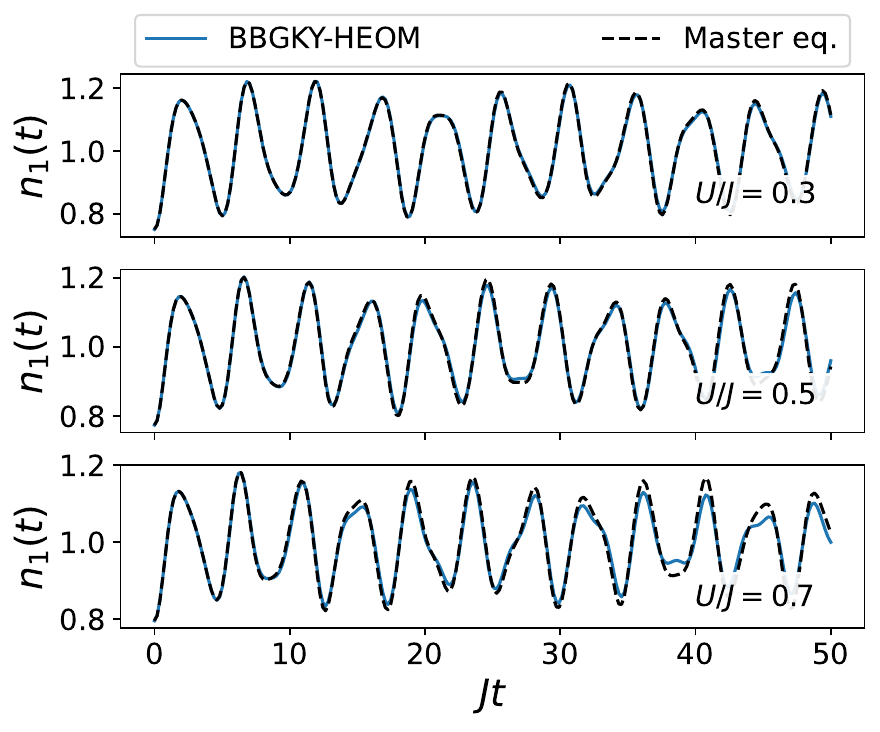}
         \end{subfigure}
         \begin{subfigure}[b]{0.52\textwidth}
             \centering
            \includegraphics[width=\textwidth]{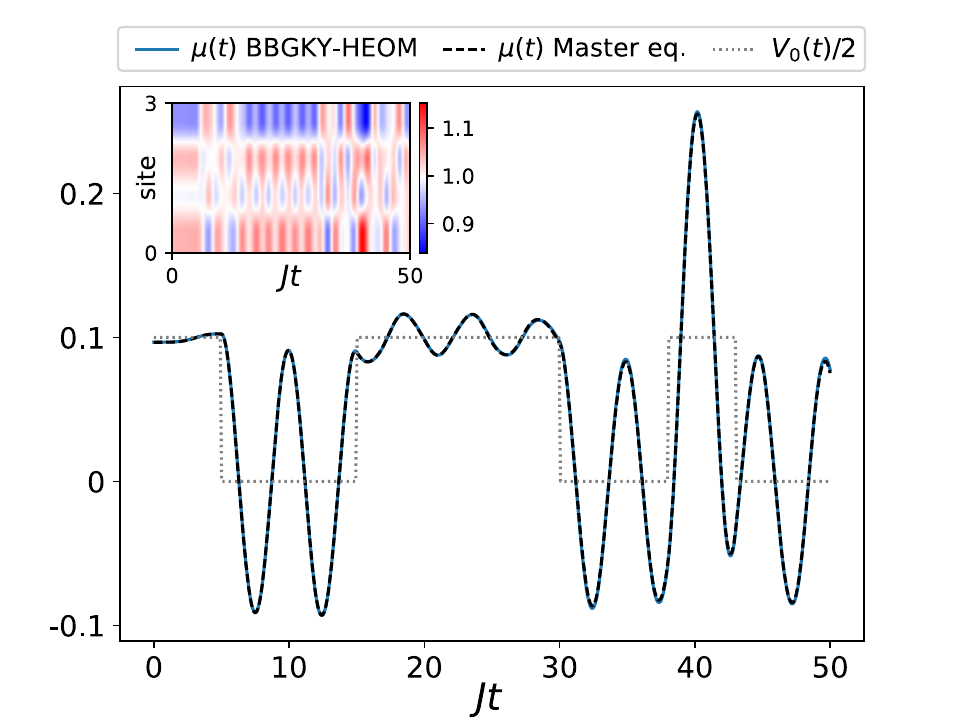}
         \end{subfigure}
    \caption{
    \textbf{Fermi-Hubbard model coupled to lossy cavity mode}:
    {(Left) Occupation of the first site as predicted by BBGKY-HEOM (solid blue) vs exact (black dashed) solution for different onside repulsion values U (see text).}
    (Right) Dipole moment driven by time-dependent potential $V_i(t) = \frac{f(t)}{5 (i+1)}$, where f(t) is a step function as shown by the grey dashed line and $U=0.1$. The example simulates charge noise due to the movement of electrons in the vicinity of the molecule. Inset illustrates onsite occupation.
    }
    \label{fig:FH_benchmark}
    \end{minipage}
\end{figure*}
We will now relax the simplification of the emitter and instead combine the dynamics of a correlated electronic system with a structured system-bath coupling.
The electronic structure of most organic molecules is dominated by covalent bonds or conjugated $\pi$-systems, requiring many-body methods to faithfully model their electronic properties.
An excitation in such a system cannot be represented by the excitation events of a single electron. 
Instead, the surrounding electronic cloud reacts and screens the polarization induced by the external field -- every optical excitation in a molecule or solid is inherently a many-body process.
Electronic correlation can reach substantial levels for e.g. transition metals, which contribute strongly localized $d$ and $f$ orbitals shielded from molecular orbitals. 
Moving two electrons to such a state costs a considerable amount of energy, as they strongly repel each other, resulting in electronic motion that is strongly correlated.
A prototypical approach to describe such a system is given by the Hubbard model

\begin{align}
\begin{split}   
    \hat{H}_{H} &= -J \sum_\sigma \sum_{i=1}^{N\add{-1}} \hat{c}_{i,\sigma}^\dagger \hat{c}_{i+1,\sigma} + \hat{c}_{i+1,\sigma}^\dagger \hat{c}_{i,\sigma}\\
    &+ U \sum_{i=1}^{N\add{-1}} \hat{c}_{i,+\frac{1}{2}}^\dagger \hat{c}_{i,-\frac{1}{2}}^\dagger \hat{c}_{i,-\frac{1}{2}} \hat{c}_{i,+\frac{1}{2}},
\end{split}
\end{align}

with the hopping rate $J$ between two localized Wannier orbitals and the onsite energy $U$. Usually $U > 0$ as it costs energy to force two electrons on the same site. Weak onsite interaction $U/J \ll 1$ will only marginally limit the free hopping of electrons between the states while extreme values of $U/J \gg 1$ will result in Mott insulators \cite{arovas2022hubbard}.
We focus here on small to moderate onsite interactions {$U/J \lesssim  1$} to account for the weak to moderate correlation in molecular systems.

Naturally, the electronic structure is embedded in an environment of optical (or vibro-phononic) modes, resulting in the emergence of polaritonic states if coupled sufficiently strongly.
Our BBGKY-HEOM approach allows to consistently describe and monitor how the dynamics of the lossy cavity mode imprints correlations in the electronic many-body system and vice versa.
The corresponding minimal Hamiltonian for such a Cavity-Fermi-Hubbard system \cite{Li2020,Mordovina2020}
\begin{align}\label{eq:hubbard}
\begin{split}
    \hat{H} &= \hat{H}_{H} + \hbar \omega (\hat{a}^\dagger \hat{a} + \frac{1}{2})\\
    &-\sqrt{\frac{\hbar\omega}{2\varepsilon_0 V}} (\hat{a}+\hat{a}^\dagger) \textbf{e}_{cav} \cdot \hat{\boldsymbol\mu}
    +\frac{1}{2\varepsilon_0 V} (\textbf{e}_{cav}\cdot \hat{\boldsymbol\mu})^2,
\end{split}
\end{align}
includes self-polarization contributions of the localized dipole moment $\hat{\boldsymbol\mu} = \textbf{e}_{chain} \sum_{\sigma}\sum_{i=1}^N r_i \hat{c}_{i,\sigma}^\dagger \hat{c}_{i,\sigma} $. The local positions $r_i \add{=\Delta (2i+1-N)/(N-1)}$ \add{($i=0,\dots,N-1$) ranging from $-\Delta$ to $\Delta$} are defined with respect to the center of the Power-Zienau-Wooley (PZW) gauge \cite{schafer2019relevance,Nazir2022}. 
{As in the previous section we incorporate cavity dissipation by means of a GKSL master equation with Lindblad operator $\sqrt{\kappa}a$.
For the numerical examples below,} we will use $\hbar\omega/J=0.5$, assign a cavity loss of $\kappa/J = 0.2$, a coupling strength of $g=\sqrt{\frac{\hbar\omega}{2\varepsilon_0 V}}\add{\Delta}=0.1J$, and disregard the self-polarization for the moment.
We emphasize that extensions to ultra-strong coupling~\cite{frisk2019ultrastrong} require the careful consideration of self-polarization terms \cite{schafer2019relevance,Nazir2022}. {The damping of the cavity is again modeled by the dissipator $\mathcal{D}[a]$ given in Eq.~\eqref{eq:dampedDrivenDicke}.}\\

One strength of explicitly treating the many-body electronic system is that the electronic response to external stimuli, such as charge noise, can be explicitly modelled.
Assume a short polyacetylene molecule with 4 sites hosting 4 electrons, represented by Eq.~\eqref{eq:hubbard}, and initially in equilibrium.
A distortion in the surrounding structure suddenly traps an electron near our molecule, resulting in a repulsive Coulomb potential $V(t) = \add{\sum_\sigma}\sum_{i=0}^{3} V_i(t) c_{i\add{,\sigma}}^{\dagger} {c}_{i\add{,\sigma}}$.
{In the following, we consider the quench dynamics after the electrons has left again, meaning we set $V_i(t<0) = \frac{4}{5 (i+1)}$ and $V_i(t\geq 0) =0$.}
{The subsequent occupation of the first site $n_1(t)=\cavg{c_1^\dagger c_1}$} is presented in Fig.~\ref{fig:FH_benchmark} (left) for three different values of onsite repulsion $U$ (blue lines). It is compared against an exact solution of the master equation (black, dashed lines).
The quality of our approximation remains excellent for { $U/J\leq0.5$} and slowly worsens for systems with sizable correlations of {$U/J=0.7$}, as one would expect.
{We note that the accuracy of BBGKY will generally also depend on the initial state, which influences to which extend non-Gaussian three-particle correlations will build up. For example, we demonstrate in Appendix ~\ref{app:purification} that accurate results can also be obtained for $U/J = 1$ with a weaker, quadratic quench potential $V_i$.}

Charge will not only localize once but randomly tunnel in and out of the impurity.
The dynamics originating from the random localization of charge on the dipole moment is shown in Fig.~\ref{fig:FH_benchmark} (right) {for $U/J=0.1$ }with the corresponding changes in occupation (inset) and time-dependent (gray dotted) potential $V_0(t)$. 
{The suddenly appearing charge shortly before $Jt=40$ enhances the oscillation already present in the system. This results} in large deviations from the equilibrium state without noticeable influence on the quality of the BBGKY-HEOM predictions.
One strength of the BBGKY-HEOM is therefore to resolve the many-body electronic dynamics that originates from time-dependent modulations which allows us to simulate e.g. charge noise, dynamics screening, or triplet-triplet annihilation explicitly -- a feature that we plan to leverage further in the future.

\subsection{Many-body dynamics embedded into organic crystals}\label{sec:hubbardwildbath}

\begin{figure}[t]
    \centering
    \includegraphics[width=\columnwidth]{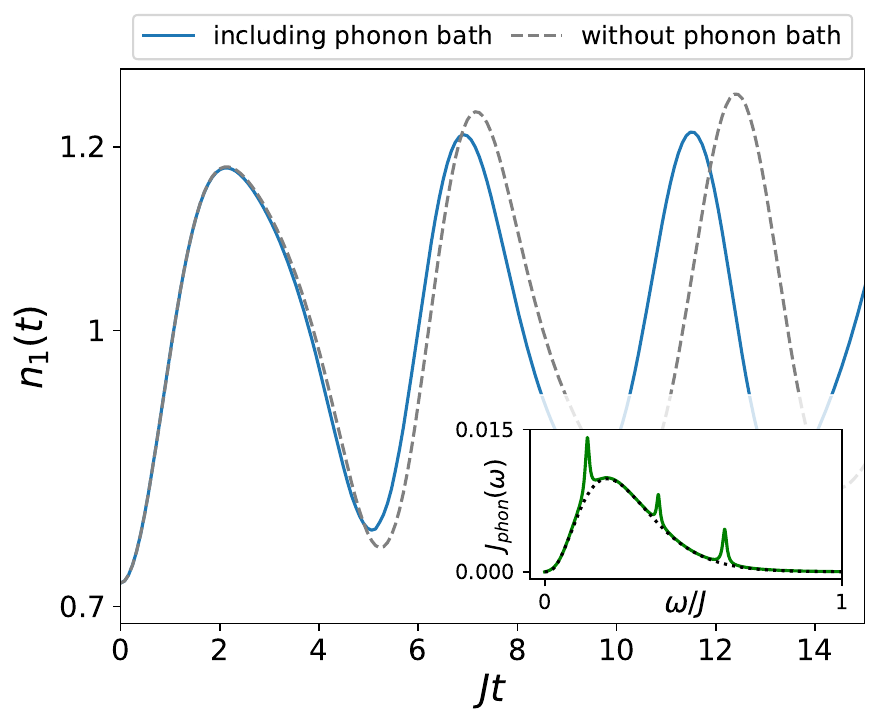}
    \caption{
    \textbf{Cavity-Fermi-Hubbard model under the influence of a complex phonon bath}:
    Occupation of the first site over time with (blue, solid) and without (gray, dashed) the phonon bath after a quench induced by charge noise. Parameters are identical to the Fermi Hubbard example in the previous section with $U=0.1$. The inset shows the phonon spectral density (green), where the superohmic background is highlighted as a black dotted line.}
    \label{fig:complexBath}
\end{figure}

Many-body electronic systems, such as molecules or (an)organic frameworks, are often embedded in a liquid or crystalline environment. %
For example, organic molecules embedded in organic crystals can serve as a highly coherent emitter~\cite{toninelli2021single,ohmanPredictive}, quantum memory~\cite{doi:10.1126/science.abb9352}, or organic light-emitting diode.
Their environment is a combination of local vibrational, extended phononic, and photonic  modes, giving rise to a highly structured and complex spectral function.
BBGKY-HEOM provides now the means to retain the complexity of the electronic structure and combine it with the full system-bath dynamic, i.e., extending beyond previous approaches that relied on a simplified description of the electronic system \cite{McCutcheon2020}.

Fig.~\ref{fig:complexBath} demonstrates the impact of a non-Markovian phonon-bath on the dynamics of the same 4-site Cavity-Fermi-Hubbard model ($U=0.1$) subsequent to a quench induced by a charge fluctuation $V_i(t\leq 0) = \frac{4}{5 (i+1)}$, $V_i(t>0)=0$. 
This model can be seen as a simplified tight-binding description of a tetracene molecule embedded in e.g. anthracene or benzene organic crystals.
As the nuclei move closer together (further apart) their effective hopping rate increases (decreases). In the context of Eq.~\eqref{eq:Htot} we model this by coupling the phonon-bath to the hopping operator \cite{PhysRevLett.42.1698}: 
\begin{equation}
    L^{phon} = \sum_\sigma \sum_{i=1}^N \hat{c}_{i,\sigma}^\dagger \hat{c}_{i+1,\sigma} + \hat{c}_{i+1,\sigma}^\dagger \hat{c}_{i,\sigma}.
\end{equation}
The corresponding phonon spectral density $J_{phon}(\omega)$ is illustrated in the inset, comprising a superohmic background as well as additional discrete modes. The superohmic background ($\propto \omega^3$) provides a simplified model for the acoustic phonon branch of a 3 dimensional crystal \cite{Mahan2000}. The total phonon coupling strength was chosen as $1/\sqrt{3}$ times the cavity coupling strength \add{($\int J_{phon}(\omega) \dd{\omega} = g^2/3$)}. Additional numerical details of the simulation can be found in Appendix \ref{app:numerics}.\\ 
Neglecting the phonon bath (gray dashed) leads to a strong misrepresentation, which is much larger than the possible deviations caused by the BBGKY approximation (recall Fig.~\ref{fig:FH_benchmark}), and clearly shows why a holistic approach to the many-body dynamics of the system and the system-bath interaction is crucial for a predictive description.
In combination with machine-learning approaches and established ab initio theory, this might provide a path towards truly predictive calculations for organic quantum light-matter interfaces.\cite{ohmanPredictive}

\section{Conclusion}\label{sec:conclusion}

No quantum system is perfectly isolated, and the description of its environment can be essential for our physical understanding.
Confined optical or plasmonic modes and vibrational or phononic modes might even challenge the dynamics of a subsystem of interest by strongly coupling to the internal degrees of freedom.
Energy deposited in the environment will then not simply decohere but act back on the system, resulting in a non-Markovian, or memory dependent, system-bath interaction.
In order to afford a rigorous treatment of such effects, the dynamics of the system itself are often simplified to the degree that the many-body character inherited by every molecule or solid-state impurity is lost.\\

In this study, we presented a rigorous hierarchical approach based on stacking of the reduced density-matrix (BBGKY) and system-bath (HEOM) hierarchies together.
BBGKY-HEOM is flexible enough to tackle critical quantum optical and condensed matter problems, striking a balance between high accuracy and good performance.
In particular, we demonstrated that correlated (superradiant) emission and spin squeezing, i.e., dynamics that extend beyond the capabilities of mean-field theory, are accurately captured.

Many-body electronic systems, here in the form of the Fermi-Hubbard model, are critical in modern condensed matter theory. We illustrate the interplay between correlated system dynamics and structured baths by simulating different forms of charge noise.
Again, excellent performance is observed as long as the onsite repulsion $U$ and the electronic dynamics remain moderate $U/J\lesssim 1$.
Finally, we demonstrated the major strength of BBGKY-HEOM by simulating the interplay between the vibro-phononic bath of an organic crystal coupled to the many-body dynamics of the Cavity-Fermi-Hubbard model.
Both contributions, i.e., non-Markovian system-bath coupling and many-body dynamics are equally important -- clearly stressing the need for a more holistic perspective on solid-state quantum emitters.\\

Our hierarchical approach 
can be rigorously informed from first principles -- bridging previously disconnected research fields.
In this way, BBGKY-HEOM provides a powerful tool that can be used alongside established methods (like cumulant expansions, density-functional theory, etc.) to gather a more detailed and holistic understanding.
Especially promising examples include molecules interacting with optical fields~\cite{kong2021probing}, may it be for quantum technological applications~\cite{ohmanPredictive,toninelli2021single,orrit}, polariton condensation~\cite{urbonas2024temporal}, or few-emitter lasing~\cite{letter_placeholder}.

\begin{acknowledgments}
We thank Walter Strunz, Alexander Eisfeld, and G\"oran Johansson for insightful discussions.
C.S. acknowledges support from the Swedish Research Council through Grant No. 2016-06059 and funding from the Horizon Europe research and innovation program of the European Union under the Marie Sk{\l}odowska-Curie grant agreement no.\ 101065117.

Partially funded by the European Union.
Views and opinions expressed are, however, those of the author(s) only and do not necessarily reflect those of the European Union or REA.
Neither the European Union nor the granting authority can be held responsible for them.
\end{acknowledgments}

\appendix

\section{Spin squeezing}\label{app:spinsqu}
The spin squeezing parameter is defined as $\xi^2(t) = N(\Delta S_{\textbf{n}_1})^2/(\cavg{S_{\textbf{n}_2}}^2 + \cavg{S_{\textbf{n}_3}}^2)$ \cite{Sorensen2001Jan},  where $\textbf{n}_1(t) \bot \cavg{\textbf{S}}(t)$ is obtained at each time point by calculating the vector perpendicular to the current spin direction that has the minimal associated variance $(\Delta S_{\textbf{n}_1})^2$. The vectors $\bvec{n}_2, \bvec{n}_3$ are then constructed such that $\{\bvec{n}_i\}$ forms a basis.

\section{Bath correlation function for a damped cavity}\label{app:cav_BCF}
\add{
In the following we provide a brief derivation of Eq.~\eqref{eq:alpha_cav} and show how the Lindblad description in Eq.~\eqref{eq:dampedDrivenDicke} can be reconciled with the total Hamiltonian \eqref{eq:Htot} given in Sec.~\ref{ssec:introManyBodyOQS}. The result is well known and a more detailed and general derivation may for example be found in the Supplemental Material of Ref.~\cite{Muller2026Apr}.}\\

\add{
For a single environment, the total Hamiltonian given in Eq.~\eqref{eq:Htot} can be written in an interaction picture with respect to the free bath Hamiltonian as
\begin{equation}\label{eq:app_Htot_B}
\begin{split}
    \Tilde{H}(t) =& H_{sys} + L^\dagger B(t) + LB^\dagger(t),\\
    B(t) =& \sum_\lambda g_\lambda^1 e^{-i\omega_\lambda^1 t}a_\lambda^1
    \end{split}
\end{equation}
The bath correlation function is obtained as $\alpha^1(\tau) = \cavg{B(\tau)B^\dagger(0)}$. In the following we will first bring the damped cavity described by Eq.~\eqref{eq:dampedDrivenDicke} in this form.\\
As the Lindblad Master equation \eqref{eq:dampedDrivenDicke} is obtained from eliminating a zero temperature Markovian bath (coupled to the cavity mode) we may equivalently express the dynamics in terms of an enlarged Hilbert space which explicitly contains the degrees of freedom in the Markovian bath:
\begin{equation}
\begin{split}
    H_{tot} =& H_{sys} + g\sum_i(\sigma_i^-a^\dagger + \sigma_i^+ a) + \Delta_c a^\dagger a \\
    &+ \sum_j \left(\Omega_jc_j^\dagger c_j + d_j(ac_j^\dagger + a^\dagger c_j)\right),\\
    \equiv&   H_{sys} +  g\sum_i(\sigma_i^-a^\dagger + \sigma_i^+ a) + H_{CM}.
\end{split}
\end{equation}
Here, the operators $c_j$ ($c_j^\dagger$) are the annihilation (creation) operators of the modes in the Markovian bath. 
In an interaction picture with respect to the cavity and bath Hamiltonian $H_{CM}$, and find
\begin{equation}\label{eq:app_evolution_int_pic}
\begin{split}
    \partial_t\ket{\Psi(t)} =& -i\Bigg[H_{sys} + \left(\sum_i \sigma_i^+\right) ga(t)\\
    &\quad\quad + \left(\sum_i \sigma_i^-\right) ga^\dagger (t)\Bigg]\ket{\Psi(t)}.
\end{split}
\end{equation}
Clearly, the total Hamiltonian is now in the form of Eq.~\eqref{eq:app_Htot_B} and $B(t)$ is given by the (yet to be determined) cavity annihilation operator in the interaction picture $ga(t)$. Thus the bath correlation function is obtained as $\alpha(\tau)=g^2\cavg{a(\tau)a^\dagger(0)}$. 
Due to the Markov property of the $c_\lambda$-bath the time evolution of $a(t)$ is given by the quantum Langevin equation \cite{Gardiner2004} (see also \cite{Muller2026Apr})
\begin{align} \label{eq:app_ai_dot}
   \dv{a}{t} 
    =& -(i\Delta_c + \kappa)a(t) -i \xi(t),
\end{align}
where $\xi(t) = \sum_j d_j e^{-i\Omega_j t} c_j$ is operator white noise with $[\xi(t),\xi^\dagger(s)] = 2\kappa\delta(t-s)$.
Thus $a(t)$ can be given explicitly as
\begin{equation}
    \label{eq:app_a_int}
    a(t) = e^{-(i\Delta_c+\kappa)t}a - i\int_0^t\dd{s}e^{-(i\Delta_c +\kappa)(t-s)}\xi(s).
\end{equation}
With the explicit solution at hand the bath correlation function is found as
\begin{align}
    \cavg{B(t)B^\dagger(s)}
        =&g^2  \cavg{a(t)a^{\dagger}(s)},\notag\\
        =& g^2e^{-i\Delta_c(t-s)-\kappa\abs{t-s}}.
\end{align}
}

\section{Derivation of the HEOM}\label{app:HEOM}
We sketch here a derivation of the HEOM equations. We assume that the system
is coupled to the environment via operator $\hat q$, such that in an interaction picture with respect to the free Hamiltonian of the environment we have
\begin{equation}
\begin{split}
    H_{tot} =& H_{sys}+\hat{q}\left(B(t) + B^\dagger(t)\right),\\
    B(t) =& \sum_\lambda g_\lambda e^{-i\omega_\lambda t}a_\lambda.
\end{split}
\end{equation}
The response of the system to the 
environment is given by the bath correlation function $\alpha(t-s) = \cavg{B(t)B^\dagger(s)}$. For HEOM we need that the bath correlation function is of the form
\begin{align}
    \alpha(t) = \sum_{j=1}^{N_{exp}} G_je^{-W_jt},
\end{align}
where $G_j$ and $W_j$ may be complex numbers. For increased readability we perform the next steps for $N_{exp} = 1$, but the extension to multiple exponentials is straight forward. The reduced dynamics of the system can be written in the path integral form 
\begin{align}
    \rho_t(q,q') &= \int\diff q_0\int\diff q_0'\int\mathcal{D}q\int\mathcal{D}q'\,\notag\\
    &\times e^{i(S[q]-S[q'])}F[q,q']\rho_0(q_0,q_0'),
\end{align}
where $S[q]$ is the action of the closed system, 
$\rho_t(q,q') = \bra{q}\rho_t\ket{q'}$ and $\hat q\vert q'\rangle = q\vert q'\rangle$ are the eigenstates of the coupling operator $\hat q$. The influence functional
contains all information about the coupling of the open system to the environment. In this case it reads 
\begin{align}
    F[q,q'] = e^{-\int_0^t\diff s\int_0^s\diff s'\,
    (q_s-q'_s)\left(\alpha(s-s')q_{s'}-\alpha^*(s-s')q'_{s'}\right)}.
\end{align}
We define auxiliary states
\begin{align}
    \rho^{(n,m)}(q,q',t) =& \int\diff q_0\int\diff q_0'\int\mathcal{D}q\int\mathcal{D}q'\,\notag\\
    &\times e^{i(S[q]-S[q'])}F[q,q']\rho_0(q_0,q_0)Q_t^n Q_t'^m,
\end{align}
where $Q_t = \int_0^t\diff s\, \alpha(t-s)q_s$ and 
$Q_t' = \int_0^t\diff s\, \alpha^*(t-s)q'_s$. Then by directly computing the 
time derivative of the auxiliary state and by using the exact expression for the open system dynamics together with the special form of the bath correlation function we directly arrive to 
\begin{align}
    \dot\rho^{(n,m)}(t) =& -i[H,\rho^{(n,m)}]-(nW+mW^*)\rho^{(n,m)}\notag\\ 
    &+[\rho^{(n+1,m)},q]+[q,\rho^{(n,m+1)}]\notag\\ 
     &+n G q\rho_t^{(n-1,m)}+mG^*\rho^{(n,m-1)}q.
\end{align}
The equations presented in the main text are a generalization of the above equations in two respects. First, we allow for multiple exponential terms in the expansion for the bath correlation function, and second we allow for the case 
$q\neq q^\dagger$ \cite{Tanimura2020}.
Additionally, it can be shown that 
\begin{equation}\label{eq:aux_states}
    \rho^{(n,m)} = i^n(-i)^m\operatorname{Tr}_{env}\left((B^\dagger(t))^mB(t)^n\rho_{tot}\right).
\end{equation}

\section{Derivation of the BBGKY-HEOM equations}\label{app:derivation}
In the following we detail the derivation of the BBGKY-HEOM equation. For the sake of readability we again focus on the case of a scalar index, corresponding to a single (global) bath that all particles couple to and a single exponential in the bath correlation function. The situation where each particle in the many-body system couples to its own (local) bath is explained in Appendix~\ref{app:localBaths}. The extension to a vector index is straight forward.
We start from the HEOM equation \eqref{eq:HEOM} and want to employ the BBGKY hierarchy to reduce the dimension of the system. \\
Following \eqref{eq:RDM} we define the reduced 2-body auxiliary matrices as
\begin{equation}
    F_{12}^{(n,m)} = N(N-1)\operatorname{Tr}_{3..N}(\rho^{(n,m)}).
\end{equation}
Their evolution equation can be derived from the HEOM Eq.~\eqref{eq:HEOM} (with $H_{sys}$ given by Eq.~\eqref{eq:Hsys}) by tracing over all but $2$ particles and assuming that the density matrix is invariant under particle exchange:
\begin{equation}
    \begin{split}
        \dot{F}_{12}^{(n,m)} =& N(N-1)\operatorname{Tr}_{3..N}(\dot{\rho}^{(n,m)}),\\
            =& -i[H_1 + H_2 + V_{12}, F_{12}^{(n,m)}] \\
            & -i \tr_3\left([V_{13} + V_{23}, F_{123}^{(\bvec{n},\bvec{m})}]\right)\\
            &- \left[(n-m)i\Delta + (m+n)\kappa\right]F_{12}^{(n,m)}\\
        &+Gn\left((L_1 + L_2)F_{12}^{(n-1,m)} + \operatorname{Tr}_3(L_3 F_{123}^{(n-1,m)})\right) \\
        &+G^*m\left(F_{12}^{(n,m-1)}(L_1^{\dagger}+L_2^{\dagger}) + \operatorname{Tr}_3(L_3^\dagger F_{123}^{(n,m-1)})\right)\\
        & +\left[F_{12}^{(n+1,m)}, L_1^{\dagger}+L_2^\dagger\right] + \left[L_1+L_2, F_{12}^{(n,m+1)}\right].
    \end{split}
\end{equation}
Above we used that operators which are traced over can be cyclically moved under the trace, such that
\begin{equation}
    \begin{split}
        \operatorname{Tr}_{3...N}\left([\sum_{i=1}^N H_i, \rho]\right) =& [H_1 + H_2, \rho_{12}] \\
        &+ \sum_{i=3}^N \operatorname{Tr}_{3...N}(H_i\rho) - \operatorname{Tr}_{3...N}(\rho H_i),\\
        =& [H_1 + H_2, \rho_{12}].
    \end{split}
\end{equation}
In addition, by utilizing the particle exchange symmetry of $\rho^{(n,m)}$, partial traces over operators can be combined with the help of the particle exchange operator $P_{3j}$ for arbitrary $j\in [4,N]$:
\begin{equation}
    \begin{split}
         \operatorname{Tr}_{3..N}(L_j \rho^{(m-1,p)}) =& \operatorname{Tr}_{3..N}( P_{3j}^2 L_j P_{3j}^2 \rho^{(m-1,p)}), \\
            =& \operatorname{Tr}_{3..N}(L_3 P_{3j} \rho^{(m-1,p)} P_{3j} ), \\
            =& \operatorname{Tr}_{3..N}(L_3 \rho^{(m-1,p)}), \\
            =& \frac{\operatorname{Tr}_{3}(L_3 F_{123}^{(m-1,p)})}{N(N-1)(N-2)}.
    \end{split}
\end{equation}
In the next step we use a cluster expansion to express the term $F_{123}^{(n,m)}$ in terms of the reduced density one- and two-body matrices. The key approximation that enables this is the neglect of higher order correlations. Combing Eq.~\eqref{eq:aux_states} and Eq.~\eqref{eq:RDM} we can express the 3-body density matrix as
\begin{equation}
    \begin{split}
        F_{123}^{(n,m)} =& N(N-1)(N-2)i^n(-i)^m\\
&\times\operatorname{Tr}_{4..N,env}\left((B^\dagger(t))^mB(t)^n\rho_{tot}\right).
    \end{split}
\end{equation}
The cluster expansion amounts to expressing the 3-body density matrix including the environment as:
\begin{equation}
\begin{split}
    \rho_{123E} =& \rho_{1}\otimes\rho_{2}\otimes\rho_{3}\otimes\rho_{E} \\
    &+ \rho_{1}\otimes\rho_{2}\otimes\Gamma_{3E} + \rho_{1}\otimes\Gamma_{23}\otimes\rho_{E} + \Gamma_{12}\otimes\rho_{3}\otimes\rho_{E} \\
    &+ \Gamma_{13}\otimes\rho_{2}\otimes\rho_{E} + \Gamma_{1E}\otimes\rho_{2}\otimes\rho_{3} + \rho_{1}\otimes\Gamma_{2E}\otimes\rho_{3}\\
    &+ \Gamma_{12}\otimes\Gamma_{3E} + \Gamma_{1E}\otimes\Gamma_{23} + \Gamma_{13}\otimes\Gamma_{2E}\\
    &+ \rho_{1}\otimes\Gamma_{23E} + \Gamma_{13E}\otimes\rho_{2} + \Gamma_{12E}\otimes\rho_{3} + \Gamma_{123}\otimes\rho_{E}\\
    &+ \Gamma_{123E}
\end{split}
\end{equation}
The operators $\Gamma_{ij}$ are defined in analogy to the 2nd cumulants as $\Gamma_{ij} = \rho_{ij} - \rho_i\otimes\rho_j$. The operators with three or four indices are defined in a similar way, using the third or fourth order cumulants and they vanish exactly for Gaussian states according to Wicks theorem. The cluster expansion can thus be understood in the sense of a cumulant expansion of the inter-particle correlations \cite{Mazziotti1998Jan}.
We now neglect the non-Gaussian 3-body correlations within the system $\Gamma_{123}=0$ as well as four body correlations $\Gamma_{123E}=0$ to obtain
\begin{equation}\label{eq:F123App}
    \begin{split}
        \widetilde{F}^{({n}, {m})}_{123} &= 4\frac{(N-1)(N-2)}{N^3}\tr(F_1^{({n},{m})}) F_1 F_2 F_3\\
        &+\frac{N-2}{N}\left(F_{12} F_3^{({n},{m})} + F_2^{({n},{m})}F_{13} + F_1^{({n},{m})} F_{23}\right) \\
        &+ \frac{N-2}{N}\left(F_{1} F_{23}^{({n},{m})} + F_{13}^{({n},{m})}F_{2} + F_{12}^{({n},{m})} F_{3}\right) \\
        &- \frac{N-2}{N^2}\tr(F_1^{({n},{m})})\left(F_{12}F_3 + F_{13}F_2 + F_1F_{23}\right)\\
        &-2\frac{(N-2)(N-1)}{N^2}\Big(F_1^{({n},{m})} F_2F_3 \\
        &\quad+ F_1F_2^{({n},{m})}F_3 + F_1F_2F_3^{({n},{m})}\Big).
    \end{split}
\end{equation}
The error of this approximation is expected to be small if (i)  the interactions between the particles are comparably weak, (ii) each particle or molecule experiences strong local dissipation, or (iii) the system is expected to behave approximately Gaussian for other reasons, e.g. due to the presence of all-to-all interactions.
This now leaves us with a reconstruction functional for all auxiliary matrices and enables us to truncate the BBGKY-HEOM equation by neglecting higher order correlations.\\

However, Eq.~\eqref{eq:F123App} does not yet preserve (anti-) symmetrization. To ensure this we need to project $\widetilde{F}^{({n}, {m})}_{123}$ onto the (anti-) symmetric subspace with the help of the projector $\Lambda_{123}^\pm = (1 +\epsilon P_{12})(1 +\epsilon P_{13} +\epsilon P_{23})$ and subsequently normalize. 
Additionally, we use the methods suggested in Refs.~\cite{Lackner,Lackner2015Feb,PhysRevResearch.5.033022} to ensure (approximate) positivity of the density matrix and contraction consistency during the dynamics. These methods are described in detail in Appendix~\ref{app:purification}.

\section{Local baths}\label{app:localBaths}
In this section we show how BBGKY-HEOM can be used to treat situations, where each particle couples to its own local bath. For the sake of readability we consider the case where the bath correlation function of each local bath is a single exponential, where, again, the generalization to multiple exponential terms is straight forward.
We start out with the HEOM equation for the full N-particle state, where the coupling operators $L_k = S_k$ only act on particle $k$ and $H_{sys} = \sum_{i=1}^N H_{sys}^{i}$ \add{(interactions can be included exactly the same way as in App.~\ref{app:derivation} without changing the argument presented below)}
\begin{equation}
    \begin{split}
        \dot{\rho}^{(\bvec{n},\bvec{m})}&=-i[H_{sys}, \rho^{(\bvec{n},\bvec{m})}] - \left(\bvec{w}\cdot\bvec{n}+\bvec{w}^*\cdot\bvec{m}\right)\rho^{(\bvec{n},\bvec{m})}\\
        &+\sum_{k=1}^N\bigg(G_kn_kS_k\rho^{(\bvec{n}-\bvec{e}_k,\bvec{m})} + G_k^*m_k\rho^{(\bvec{n},\bvec{m}-\bvec{e}_k)}S_k^{\dagger}\\
        &+\left[\rho^{(\bvec{n}+\bvec{e}_k,\bvec{m})}, S_k^{\dagger}\right] + \left[S_k, \rho^{(\bvec{n},\bvec{m}+\bvec{e}_k)}\right]\bigg),
    \end{split}
\end{equation}
Now after tracing out particle $N$, with $\rho_{1..N-1}^{(\bvec{n},\bvec{m})} = \operatorname{Tr}_N(\rho^{(\bvec{n},\bvec{m})})$ we are left with
\begin{equation}\label{eq:HEOM_lb_2}
    \begin{split}
        \dot{\rho}_{1..N-1}^{(\bvec{n},\bvec{m})})=&-i[\sum_{i=1}^{N-1}H_{sys}^{i}, \rho_{1..N-1}^{(\bvec{n},\bvec{m})}] - \left(\bvec{w}\cdot\bvec{n}+\bvec{w}^*\cdot\bvec{m}\right)\rho_{1..N-1}^{(\bvec{n},\bvec{m})}\\
        &+\sum_{k=1}^{N-1}\bigg(G_kn_kS_k\rho^{(\bvec{n}-\bvec{e}_k,\bvec{m})}_{1..N-1} + G_k^*m_k\rho_{1..N-1}^{(\bvec{n},\bvec{m}-\bvec{e}_k)}S_k^{\dagger}\\
        &\quad\quad\quad+\left[\rho_{1..N-1}^{(\bvec{n}+\bvec{e}_k,\bvec{m})}, S_k^{\dagger}\right] + \left[S_k, \rho_{1..N-1}^{(\bvec{n},\bvec{m}+\bvec{e}_k)}\right]\bigg)\\
        &+G_Nn_N\operatorname{Tr}_N(S_N\rho^{(\bvec{n}-\bvec{e}_N,\bvec{m})}) \\
        &+ G_N^*m_N\operatorname{Tr}_N(\rho^{(\bvec{n},\bvec{m}-\bvec{e}_N)}S_N^{\dagger})\Big).
    \end{split}
\end{equation}
We see that the last two terms actually still depends on hierarchy states with $n_N, m_N > 0$. However, ultimately we are only interested in the physical density matrix  $\operatorname{Tr}_N(\dot{\rho}^{(\bvec{0},\bvec{0})})$. We can see that the evolution of this physical density matrix only depends on auxiliary states with $m_N = n_N = 0$, because the sum in line two and three only extends to $N-1$. Furthermore, the evolution of all auxiliary states with $m_N = n_N = 0$ do in turn also not depend on auxiliary states with $m_N,n_N > 0$. Thus the physical density matrix is completely independent of this part of the hierarchy and it may as well be omitted. After omitting the last line in Eq.~\eqref{eq:HEOM_lb_2} we have thus eliminated the particle $N$ together with its local bath. Continuing to trace out all particles except 1 and 2 we are then left with
\begin{equation}\label{eq:HEOM_lb}
    \begin{split}
        \dot{F}_{12}^{(\bvec{n},\bvec{m})}=&-i[H_{sys}^{1} + H_{sys}^{2}, F_{12}^{(\bvec{n},\bvec{m})}] \\
        &- \left(\bvec{w}\cdot\bvec{n}+\bvec{w}^*\cdot\bvec{m}\right)F_{12}^{(\bvec{n},\bvec{m})}\\
        &+\sum_{k=1}^{2}\bigg(G_kn_kL_k F_{12}^{(\bvec{n}-\bvec{e}_k,\bvec{m})} + G_k^*m_k F_{12}^{(\bvec{n},\bvec{m}-\bvec{e}_k)}L_k^{\dagger}\\
        &+\left[F_{12}^{(\bvec{n}+\bvec{e}_k,\bvec{m})}, L_k^{\dagger}\right] + \left[L_k, F_{12}^{(\bvec{n},\bvec{m}+\bvec{e}_k)}\right]\bigg).
    \end{split}
\end{equation}

\section{Purification and contraction consistency}\label{app:purification}
For fermionic systems one straight-forward approach to obtain the (antisymmetric) 3-RDM is to project the reconstructed 3-RDM from Eq.~\eqref{eq:F123App} onto the antisymmetric manifold as
\begin{equation}
    \!^RF_{123}^{(n,m)} = \Lambda_{123}^- \widetilde{F}_{123}^{(n,m)}\Lambda_{123}^-/\mathcal{N},
\end{equation}
where $\Lambda_{123}^- = \Lambda_{12}^-(1 - P_{13} - P_{23})$, $\Lambda_{12}^-=1-P_{12}$, $P_{ij}$ is the particle exchange operator for particles $i,\, j$ and $\mathcal{N}$ is a normalization. However, $\!^RF_{123}^{(0,0)}$ will generally not represent a physical reduced state of an anti-symmetric $N$-particle density matrix \add{, an issue known as the N-representability problem} \cite{N_representability_conditions}. 
Furthermore, we find that $\!^RF_{123}^{(n,m)}$ is not contraction consistent, meaning that
\begin{equation}\label{eq:contractionConsistencyViolation}
    \operatorname{Tr}_3(\,\!^RF_{123}^{(n,m)}) \neq (N-2)F_{12}^{(n,m)}.
\end{equation}
These inconsistencies will generally lead to unphysical artifacts, such as negative eigenvalues of $F_{12}^{(0,0)}(t)$, and instabilities in the propagation.
However, it was shown in Refs.\cite{Lackner,Lackner2015Feb,PhysRevResearch.5.033022} that these instabilities can be cured to a large extent by (1) enforcing contraction consistency of the 3-RDM and (2) using a so called dynamical purification scheme, that enforces the positivity of the 2-RDM as well as the two hole RDM.
Below we describe our implementation that enforces these two conditions for the simulations involving fermions, following \cite{Lackner,Lackner2015Feb}.\\

\begin{figure*}
    \includegraphics[width=0.45\textwidth]{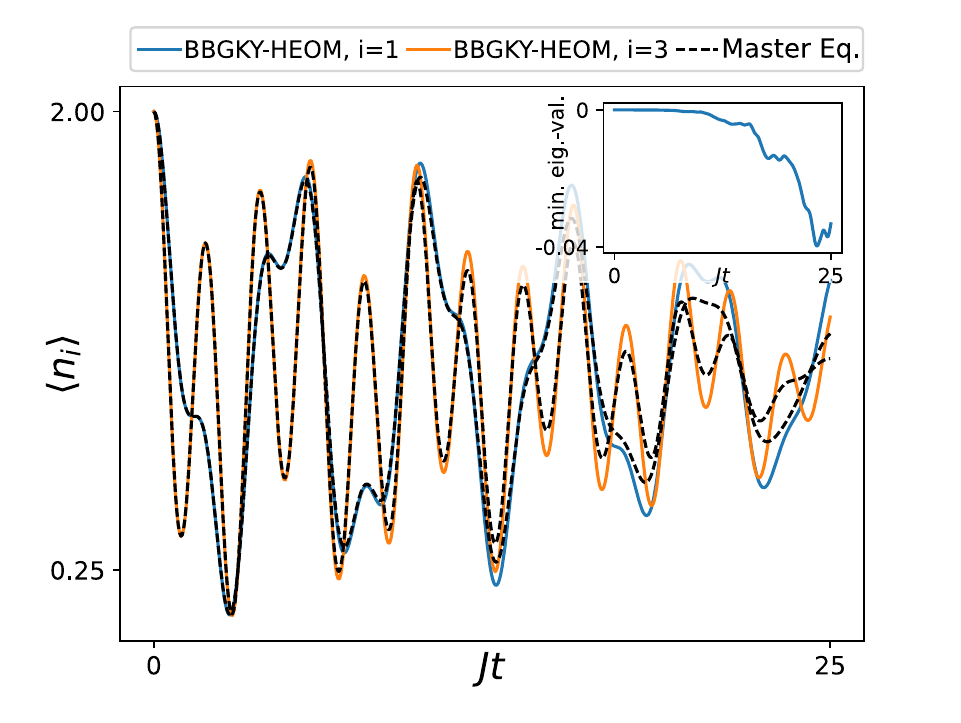}
    \includegraphics[width=0.45\textwidth]{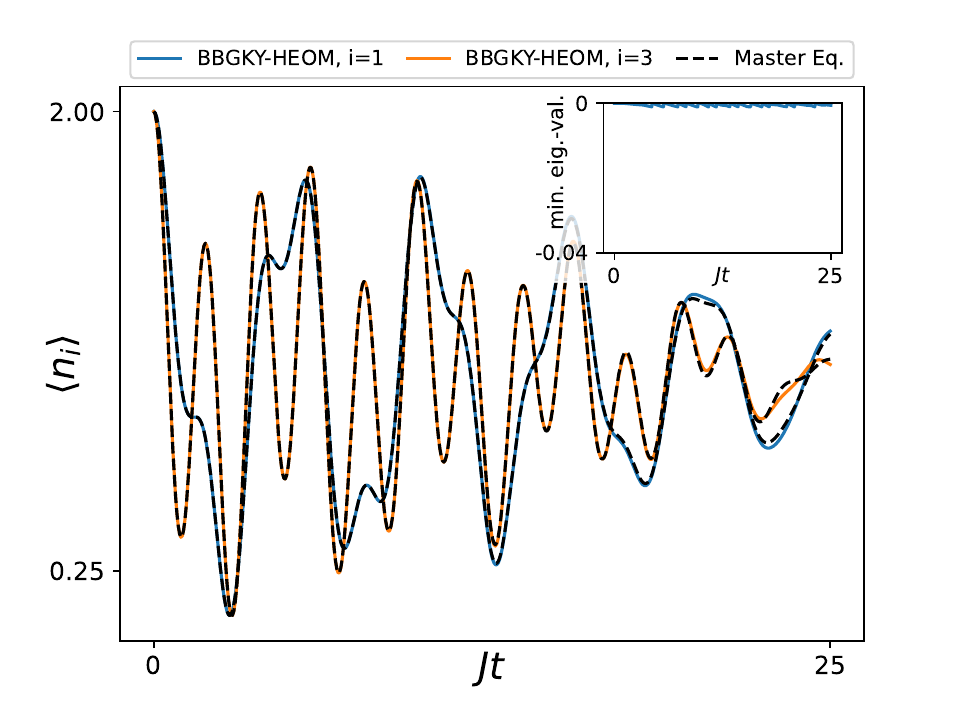}
    \caption{Simulation of a four site Hubbard chain with $U/J = 0.1$, where initially the sites one and three are doubly occupied and the sites 2 and 4 are vacant. The right (left) plot shows the simulation with (without) a contraction consitent reconstruction and purification. Insets show the minimal eigenvalue of $F_{12}(t)$ over time.}
    \label{fig:purificationDemonstration}
\end{figure*}
We use the notation $\rho_{j_1,j_2,\dots,j_N}^{i_1,i_2,\dots,i_N}$ for $N$-particle density matrices $\rho_{(i_1,i_2,\dots,i_N),(j_1,j_2,\dots,j_N)}$, where each index runs from one up to the single particle dimension.
A key concept for the implementation below is the unitary decomposition of matrices \cite{unitaryDecpmosition1,unitaryDecpmosition2}, that allows to uniquely decompose every matrix $M$ as
\begin{equation}
    M =\!_0 M + \!_\perp M,
\end{equation}
where every index contraction (and thus in particular every partial trace) of $\,\!_0M_{j_1,j_2,\dots,j_N}^{i_1,i_2,\dots,i_N}$ vanishes and $\!_0 M,\, \!_\perp M$ are orthogonal $\operatorname{Tr}\left(\!_0 M^\dagger \, \!_\perp M\right) = 0$.
For antisymmetric two and three-particle density matrices $M_{12},\,M_{123}$ with single particle dimensions $d_1$ we find the following explicit expressions for the orthogonal component
\begin{align}
    \!_\perp M_{12} =& \frac{\Lambda_{12}^-(M_1\otimes\id)\Lambda_{12}^-}{d_1-2} - \frac{\Lambda^-_{12}\tr{(M_1)}}{(d_1-2)(d_1-1)},\label{eq:kernelM12}\\
    \!_\perp M_{123} =& \frac{\Lambda_{123}^-(M_{12}\otimes\id)\Lambda_{123}^-}{4(d_1-4)} - \frac{\Lambda^-_{123}(M_1\otimes\id\otimes\id)\Lambda^-_{123}}{2(d_1-4)(d_1-3)}\label{eq:kernM123}\\
    &+\frac{\tr(M_1)\Lambda_{123}^-}{(d_1-4)(d_1-3)(d_1-2)}.\notag
\end{align}
A crucial observation is that the orthogonal components $\!_\perp M_{123},\,\!_\perp M_{12}$ can be constructed from the knowledge of the corresponding reduced states $M_1=\tr_2(M_{12}),\,M_{12}=\tr_3(M_{123})$ alone: $\!_\perp M_{123} = \!_\perp M_{123}\left(M_{12},M_1\right)$. \\

We can use the unitary decomposition to enforce the construction consistency of the (antisymmetric) 3RDM. For this, we first calculate the mismatch on the level of the reduced state
\begin{equation}
    \!^\Delta F_{12}^{(n,m)} \equiv F_{12}^{(n,m)} - \frac{1}{N-2}\operatorname{Tr}_3\left(\!^RF_{123}^{(n,m)}\right).
\end{equation}
With $\!^\Delta F_{1}^{(n,m)} = \operatorname{Tr}_2(\,\!^\Delta F_{12}^{(n,m)})$ we then obtain the corresponding orthogonal component $\!^\Delta_\perp F_{123}^{(n,m)}\left(\!^\Delta F_{12}^{(n,m)}, \!^\Delta F_{1}^{(n,m)}\right)$ according to Eq.~\eqref{eq:kernM123}. Following Ref.~\cite{Lackner2015Feb} the contraction consistent, antisymmetric 3-RDM $\!^{AS} F_{123}^{(n,m)}$ is then given as
\begin{equation}\label{eq:antiSymmetricReconstruction}
    \!^{AS} F_{123}^{(n,m)} = \!^RF_{123}^{(n,m)} + (N-2)\cdot\!^\Delta_\perp F_{123}^{(n,m)}.
\end{equation}
For all simulations involving fermions we use $\!^{AS} F_{123}^{(n,m)}$ to approximate the three-particle state in Eq.~\eqref{eq:reducedHEOM}, unless stated otherwise.\\

In addition, we employ a dynamical purification scheme \cite{Lackner2015Feb} for the fermionic simulations, which ensures that the smallest eigenvalue of the \add{(normalized)} reduced 2-particle density  $F^{(0,0)}_{12}(t)\add{/\operatorname{Tr}(F^{(0,0)}_{12}(t))}$ never goes below $-\epsilon_p$, where the threshold $\epsilon_p$ was chosen as $\epsilon_p = 10^{-3}$ for the simulations in this manuscript.
To ensure this, we track the smallest eigenvalue of $F^{(0,0)}_{12}(t)$ during runtime and perform a series of purification steps once the smallest eigenvalue falls below threshold. 
These purification steps were implemented following Ref.~\cite{Lackner2015Feb} and amount to the following procedure:
\begin{enumerate}
    \item{We construct the two hole RDM $Q_{12} = \Lambda_{12}^-/2 - \Lambda_{12}^-\left(\id\otimes F_1\right)\Lambda_{12}^-/4 + F_{12}$, where $\Lambda_{12}^- = \id-P_{12}$.}
    \item{We obtain the eigenvalues $\lambda_i^F$ ($\lambda_i^Q$) and eigenvectors $\ket{v_i^F}$ ($|{v_i^Q}\rangle$) of $F_{12}$ ($Q_{12}$) and decompose them into their positive and negative parts.
    \begin{equation}
        F_{12}^< = \sum_{\lambda_i^F<0}\lambda_i^F\ketbra{v_i^F}{v_i^F},\quad Q_{12}^< = \sum_{\lambda_i^Q<0}\lambda_i^Q\ketbra{v_i^Q}{v_i^Q}.
    \end{equation}}
    \item{We compute the unitary decomposition of $F_{12}^<,\,Q_{12}^<$ according to Eq.~\eqref{eq:kernelM12} and subtract the kernel components $\!_0F_{12}^<,\,\!_0Q_{12}^<$ from $F_{12}$ to obtain a purified state $\!^pF_{12} = F_{12} - \!_0F_{12}^< -\!_0Q_{12}^<$. $\!^pF_{12}$ will have significantly reduced negative eigenvalues \cite{Lackner2015Feb}.
    }
    \item{If the smallest eigenvalue of  $\!^pF_{12}/\operatorname{Tr}(\,\!^pF_{12})$ is above a certain threshold (chosen as $-10^{-5}$ for the simulations in this manuscript) we accept $\!^pF_{12}$ as the new 2RDM and continue the calculation. If not, we repeat the purification steps.}
\end{enumerate}
\add{This purification scheme imposes two (out of many \cite{N_representability_conditions}) necessary N-representability conditions, and thus brings $F_{12}$ closer to a physically valid (fermionic/bosonic) 2-RDM \cite{Lackner2015Feb}.}\\
The contraction consistent reconstruction in Eq.~\eqref{eq:antiSymmetricReconstruction} and the purification significantly extend the range of accessible parameters as well as the accuracy of the method. An example of this is shown in Fig.~\ref{fig:purificationDemonstration}, where we compare compare a simulation without (left plot) and with (right plot) the contraction consistent reconstruction and purification.
The simulation shows the dynamics of the cavity coupled four-site Hubbard chain discussed in Sec.~\ref{sec:hubbard}, with $U/J=0.1$.
The 4 sites of our Hubbard-chain host again 4 electrons but we force them initially to  occupy the 1st and 3rd site only, i.e., our initial state has two doubly occupied and two empty sites. 
Naturally, this state is far from the correlated ground-state and results in a violent dynamics of density sloshing forward and backward. 
The insets show the respective minimal eigenvalue of $F_{12}$ over time. The comparison with an exact solution of the Master equation (black dashed line) clearly shows how the contraction consistent reconstruction and purification (right plot) significantly improve the accuracy of BBGKY-HEOM and prevent unphysical states.\\

\begin{figure}
    \centering
    \includegraphics[width=\linewidth]{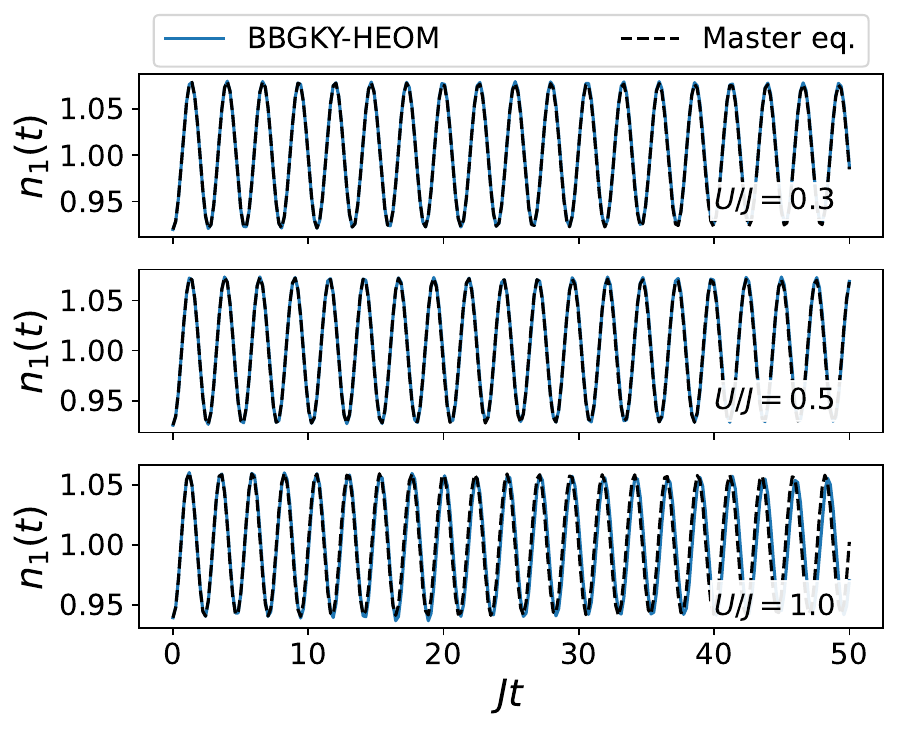}
    \caption{Comparison of BBGKY-HEOM to an exact reference calculation for different coupling strengths $U$, similar to Fig.~\ref{fig:FH_benchmark}, but with a weaker, quadratic quench potential (see text).}
    \label{fig:quadraticQuench}
\end{figure}

Additionally, we confirm in Fig.~\ref{fig:quadraticQuench} that our implementation can be applied in a parameter regime that is roughly equivalent to Ref.~\cite{Lackner2015Feb}. Using a quadratic quench potential to obtain the initial state $V_i(t<0)=1/8(i-1.5)^2$, $V_i(t>0)=0$ and other parameters identical to Sec.~\ref{sec:hubbard} we find, that BBGKY-HEOM produces accurate results up to $U/J=1.0$.\\

However, as the complete hierarchy including all auxiliary states $\rho_{HEOM} = \sum_{n,m} F_{12}^{(n,m)} \otimes \ketbra{n}{m}$ is generally not guaranteed to be positive, the purification described above can only be applied on the level of $F_{12}^{(0,0)}$ instead of the whole the hierarchy. 
For particularly strongly coupled environment the purification therefore becomes less effective in preventing negative eigenvalues. 
One possible direction for future improvements would be to incorporate the recent results of Ref.~\cite{Muller2026Apr}, which suggest a fitting ansatz for the bath correlation function that \emph{does} guarantee the positivity of $\rho_{HEOM}$ and therefore allows to consistently apply the purification on the whole hierarchy.

\section{Multipolar gauge}\label{app:multi}
Electric and magnetic fields can be expressed in terms of the vector and 
scalar potentials
\begin{align}
    E &= -\nabla \phi - \dot A,\\
    B &= \nabla\times A,
\end{align}
where $\dot A :=\frac{\partial}{\partial t}A$~\cite{craig1998molecular}.
The $E$ and $B$ fields are invariant with respect to gauge transformations
\begin{align}
    A' = A-\nabla\chi,\quad \phi' = \phi+\dot\chi.
\end{align}
We define the Green's function for the divergence operator as 
\begin{align}
    \nabla\cdot g^\parallel(x,x') = -\delta(x-x').
\end{align}
We can add to $g^\parallel(x,x')$ any transversal vector field $g^\perp(x,x')$ and 
define 
\begin{align}
    g_\mu(x,x') = g^\parallel(x,x') +\mu g^\perp(x,x').
\end{align}
The parallel Greens function has the well known representation
\begin{align}
    g^\parallel(x,x') = \nabla\frac{1}{4\pi\vert x-x'\vert}.
\end{align}
A possible choice for the perpendicular Green's function is 
\begin{align}
    g^\perp(x,x') = \int_0^1\diff \lambda x'\cdot\delta^\perp(\lambda x'-x),
\end{align}
which corresponds to a straight line from the origin to the point $x'$. This particular choice is motivated by the construction of the multipolar gauge~\cite{woolley2022foundations}.
We define a gauge function $\chi_\mu$ as 
\begin{align}
    \chi_\mu(x) = \mu\int\diff x'\,g^\perp(x',x)A^\perp(x').
\end{align}
We can thus define new vector potential as 
\begin{align}
    A_\mu(x) = A^\perp(x)-\mu\nabla\chi(x).
\end{align}
We can define a polarization field also with the help of the Greens function
\begin{align}
    P_\mu(x) = \int\diff x'\, g_\mu(x,x')\rho(x'),
\end{align}
where $\rho(x)$ is the charge distribution. We see immediately that $\nabla\cdot P_\mu(x) = -\rho(x)$.
The choice $\mu=0$ corresponds to a Coulomb gauge and $\mu=1$ to the multipolar gauge~\cite{woolley2022foundations}. We may derive the Hamiltonian in arbitrary $\mu$-gauge from a Lagrangian formalism and we find that~\cite{Nazir2022}
\begin{align}
    H=&\sum_i \frac{1}{2m_i}(p_i-q A_\mu(r_i))^2+V\notag\\
    &+\frac{1}{2}\int\diff x\, 
    (\Pi_\mu(x)+P_\mu^\perp(x))^2+(\nabla\times A^\perp(x))^2),
\end{align}
where $p_i,x_i$ and $A^\perp,\Pi_\mu$ are canonically conjugate and satisfy the commutation relations $[x_{i,a},p_{j,b}] = i\hbar\delta_{ij}\delta_{ab}$
and $[A^\perp_{a}(x),\Pi_{\mu,b}(x')] =i\hbar\delta_{ab}^\perp(x-x')$.

We assume that the center of mass of the system is at the origin and do the electric dipole approximation~\cite{Nazir2022}. The Hamiltonian simplifies to 
\begin{align}
    H &= \sum_i\frac{(p_i-(1-\mu)A_\mu(0))^2}{2m}+V\\
    &+\sum_i\mu d_i\cdot\Pi_\mu(0)
    +\frac{1}{2}\mu^2\sum_i d_i\delta^\perp(0)d_i\\
    &+\frac{1}{2}\int\diff x\, (\Pi_\mu(x)^2+(\nabla\times A^\perp(x))^2).
\end{align}

In the Coulomb gauge $(\mu=0)$ we obtain that the coupling to the electromagnetic field is provided by the term $(p_i-A_\mu(0))^2$ leading to the diamagnetic $A^2$ term. 

On the other hand, in the multipolar gauge ($\mu=1$) the transition dipole of the system $d_i$ couples to the field $\Pi_1$. Physically this means that the system couples to the field which is the difference of the transverse electric field generated bu the transverse vector potential and the polarization field
\begin{align}
    \Pi_1= E^\perp - P^\perp.
\end{align}
In addition a self-polarization term arises. 

The choice of the gauge leads to different types of interpretation of the physical nature of the electric field the system couples to. We refer further discussions on this subtle points to literature~\cite{craig1998molecular,cohen1989photons,woolley2022foundations,Nazir2022}

\section{Numerical details}
\label{app:numerics}
\begin{figure}[t]
    \centering
    \includegraphics[width=0.95\linewidth]{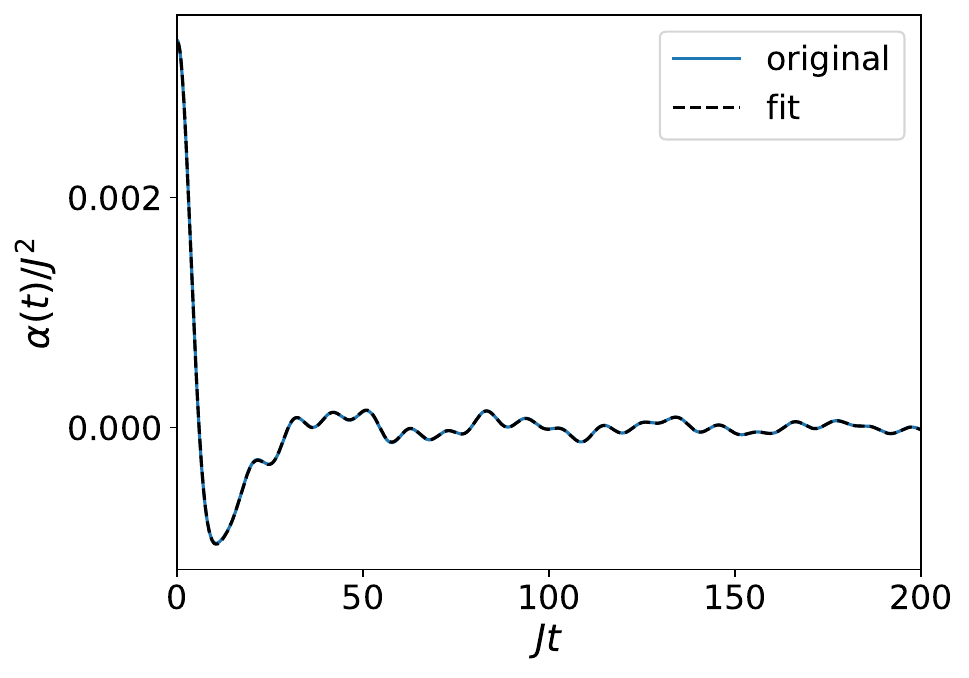}
    \caption{Fitted and exact bath correlation function for the spectral density shown in the inset of Fig.~\ref{fig:complexBath}. For the fit 6 exponential terms were used.}
    \label{fig:bathFitComparison}
\end{figure}

Numerical simulations were performed with the python packages numpy\cite{numpy}, scipy\cite{scipy} and qutip\cite{qutip}. 
The BBGKY-HEOM equation \eqref{eq:reducedHEOM} with the approximation for the three-body reduced density matrix given in Eq.~\eqref{eq:F123} were solved with the scipy implementation of the explicit Runge-Kutta method of order 5(4), with an absolute tolerance of $1e-10$ and dynamical time stepping. 

We checked for convergence with respect to the hierarchy depth, which justifies a hierarchy depth of 5 for the Tavis-Cummings model and 3 for the Cavity-Fermi-Hubbard model.
The benchmark simulations for the quantum master equation were performed using qutips mesolve function with default parameters.\\

As an example, we describe the application of BBGKY-HEOM to the system in Sec.~\ref{sec:hubbardwildbath} step by step. As motivated in the main text we use a spectral density with a few discrete peaks placed on top of a superohmic background. With the spectral density at hand we obtain the bath correlation function according to $\alpha(\tau) = \int_0^\infty J(\omega)\left(\coth{(\omega/(2T))}\cos{(\omega\tau)} -i\sin{(\omega\tau})\right) \diff{\omega}$. For our case of $T=0$ this integral can easily be performed numerically with the help of standard Fast-Fourier-Transform algorithms \cite{scipy}. 
To fit the bath correlation function with exponentials (according to Eq.~\eqref{eq:fitExp}) we simply use standard minimization algorithms \cite{scipy} to minimize the $\mathrm{L}^1$ distance between the fit and the exact function. 
However, more sophisticated algorithms are also available \cite{expFittingOverview}. 
The resulting fit is shown in Fig.~\ref{fig:bathFitComparison}.
With the coupling operator $L$ and the fit parameters $G_k,\,W_k$ from Eq.~\eqref{eq:fitExp} at hand, we can now construct and solve the BBGKY-HEOM equations. 
Since we have a global bath coupling in our example (all electrons couple to the same vibrations) we use Eq.~\eqref{eq:reducedHEOM} together with Eq.~\eqref{eq:F123} to make predictions. For baths which are local to each particle we refer to Appendix ~\ref{app:localBaths}.

Finally, we check the validity of the solution by investigating the occurrence of significant negative eigenvalues. For the application in Sec.~\ref{sec:hubbardwildbath} we find that all eigenvalues of $F_{12}^{(0,0)}$ are $>-0.015$ at all times (while $\tr{F_{12}^{(0,0)}} = 12$). For larger $U$ and larger coupling strengths we find that more severe negative eigenvalues occur, which we attribute to the shortcomings of the purification scheme, as described in Appendix \ref{app:purification}.

\bibliography{library}

\end{document}